\documentclass[aps,english,preprint,nofootinbib]{revtex4}

\usepackage{hyperref}
\usepackage{graphicx}
\usepackage{amsmath, amssymb}
\usepackage{babel}
\usepackage{color}
\usepackage{mathrsfs}
\usepackage{slashed}
\usepackage{setspace}
\usepackage{array}
\usepackage{cancel}
\usepackage{tikz}
\usepackage{appendix}
\usepackage{amsmath, bm}

\usepackage[latin9]{inputenc}
\usepackage{array}
\usepackage{varwidth}
\usepackage{amstext}
\usepackage{titlesec} 
\usepackage{xcolor}

\usepackage[T1]{fontenc} 
\usepackage[normalem]{ulem} %
\usepackage{float}
\usepackage{epstopdf}
\usepackage{subfigure}
\usepackage{bbm}
\usepackage{makecell}
\UseRawInputEncoding
\usepackage{multirow} 
\newcommand{\authornote}

\makeatletter

\providecommand{\tabularnewline}{\\}

\begin{document}
		
\title{Matching Wilson flow time for calculating the topological charge density and the pseudoscalar glueball mass in quenched lattice QCD}

\author{Zhen Cheng$^{a}$}
\thanks{Corresponding author. \\ E-mail addresses: zjuercz@zju.edu.cn (Zhen Cheng), xionggy@zju.edu.cn (Guang-yi Xiong).}
\author{Guang-yi Xiong$^{b}$}

\affiliation{$^{a}$Department of Science Education, School of Education, Zhejiang
International Studies University, Hangzhou 310023, China \\
$^{b}$Department of Physics, School of Information Engineering, Jiangxi Science and Technology Normal University, Nanchang 330036, China}

\begin{abstract}
The matching Wilson flow time for calculating the topological charge density correlator (TCDC) of the gluonic definition by the Wilson flow is analyzed using the matching procedure. The relationship has been established between the matching Wilson flow time for calculating TCDC and the matching Wilson flow time at which the topological charge defined by the gluon fields is closest to an integer. The properties of TCDC defined by the bosonic field are investigated, and the pseudoscalar glueball mass was extracted from the TCDC computed at the matching Wilson flow time. It is further demonstrated that the matching Wilson flow time determined by the matching procedure can be applied to compute the topological susceptibility in the gluonic definition. 
\end{abstract}

\maketitle

\section{Introduction }

The QCD vacuum is believed to possess a non-trivial topological structure, characterized by the topological charge and topological charge density. These topological properties are crucial for understanding various phenomena, including the $\text{U}\left(1\right)_{A}$ problem, confinement, $\theta$ dependence, and spontaneous chiral symmetry breaking \cite{Schierholz:1994pb,Witten:1978bc,Diakonov:1995ea,Bonati:2015sqt}. Lattice QCD is a powerful tool for investigating these topological properties from first principles. Definitions of topological charge and topological charge density are generally divided into the fermionic and gluonic definitions \cite{Alexandrou:2017hqw}. On the lattice, results of the gluonic and fermionic definitions of the topological charge are consistent in the continuum limit $a\to0$ \cite{Fujikawa:1998if,Kikukawa:1998pd}.
The topological charge, defined using the fermion operator that satisfies the Ginsparg-Wilson relation, is an integer \cite{Hasenfratz:1998ri}; however, the computational cost of this method is prohibitively high. Computing the topological charge on the lattice using the gluonic definition is less computationally demanding, but the numerical value of the topological charge is typically not an integer due to ultraviolet fluctuations in the gauge fields. To ensure that the numerical value of the topological charge nears an integer, a renormalization constant must be applied, or smoothing techniques must be used. Smoothing procedures, such as cooling, smearing (including APE, HYP, and stout smearing), or gradient flow, are commonly employed in the computation of the topological charge density \cite{Alexandrou:2017hqw}. However, when calculating the topological charge or topological charge density of gluonic definition, determining the matching Wilson flow time is a question of interest. In this work, we introduce a method to determine the matching Wilson flow time for calculating the topological charge density using the gluonic definition by comparing the topological charge density computed via the SMP method (fermionic definition) with those obtained from the Wilson flow (gluonic definition).

The topological charge density correlator (TCDC) is negative at any non-zero distances due to the reflection positivity and the pseudoscalar nature of the relevant local operator in Euclidean field theory. The negativity of the TCDC has significant implications for the nature of the topological charge structure in the QCD vacuum \cite{Chowdhury:2012sq,Vicari:1999xx,Horvath:2005cv}. For instance, the pseudoscalar glueball mass can be extracted from the TCDC in pure gauge theory. The mass of the $\eta^{\prime}$ meson can be extracted from the TCDC in $N_f=2+1$ lattice QCD \cite{Fukaya:2015ara}. The correlator of gluonic observables exhibits large vacuum fluctuations, making the extraction of glueball masses significantly more challenging compared to hadronic masses. However, extracting the pseudoscalar glueball mass from the TCDC does not require calculating connected and disconnected quark diagrams \cite{Creutz:2010ec}. In lattice QCD, due to severe singularities and lattice artifacts in the TCDC, smoothing of the gauge field is necessary. It is well known that undersmearing cannot completely remove lattice artifacts, while oversmearing may wipe out even the negative nature of the correlator \cite{Chowdhury:2014mra}. The impact of the smoothing radius on the sphaleron rate defined from the TCDC had been researched \cite{kotov2018sphaleron,Altenkort:2020axj,Bonanno:2023ljc,Bonanno:2023thi}. Therefore, the degree of smoothing (Wilson flow) in the calculation of TCDC  is worthy of study.

The topological susceptibility $\chi$ can be obtained by the four-volume integral of the TCDC \cite{Smit:1986fn}. Topological susceptibility, which reflects the fluctuations of the topological charge, is of great importance in the study of the QCD vacuum. The universality of the topological susceptibility in the fermionic definition shows that it is free of short-distance singularities \cite{Luscher:2010ik}. The topological susceptibility $\chi$ is linked to the $\text{U}\left(1\right)$ anomaly and the mass of $\eta^{\prime}$ meson, as expressed in the well-known Witten-Veneziano relation  \cite{Witten:1979vv,Veneziano:1979ec,DelDebbio:2004ns,Cichy:2015jra,Ce:2015qha,Bonanno:2019xhg,Bonanno:2023ple}.

The overlap operator, as a solution to the Ginsparg-Wilson equation \cite{Neuberger:1997fp,Neuberger:1998wv}, is commonly used to calculate the topological charge of the fermionic definition. Traditionally, the topological charge density has been calculated by using the point source \cite{Horvath:2003yj,Ilgenfritz:2007xu}, which is almost impossible on a large volume lattice. To address this, Ref. \cite{Xiong:2019pmh} proposed the symmetric source (SMP) method to calculate the topological charge density  of the fermionic definition. Although the SMP method can reduce the computational resources required for calculating the topological charge, it remains challenging to apply the SMP method for calculating TCDC or the topological susceptibility in the context of when the number of configurations is large. In current practical calculations, we generally use the gluonic definition for the topological charge density to compute the topological susceptibility or TCDC. However, when calculating the topological charge density correlator of the gluonic definition, the degree of smoothing remains a question worthy of research. The matching Wilson flow time for two cases are considered in this paper, and the definitions will be given later. The first case involves determining the matching Wilson flow time $\tau_{\text{qmr}}$ for calculating the topological charge by requiring that the gluonically defined topological charge is closest to integer values. The second case is to compute the matching parameter and determine the matching Wilson flow time $\tau_{\text{mr}}$ when this parameter is closest to 1.

The optimal matching parameters for comparing the results of the SMP method under different parameters, as well as the optimal matching parameters for comparing the results of the SMP method with those of the Wilson flow at different flow times, will be calculated. By analyzing these matching parameters, the matching Wilson flow time for calculating the TCDC will be determined. Further exploration of the SMP method in the calculation of the topological charge of the fermionic definition will be presented. The relationship between the TCDC and topological susceptibility with the matching parameters will be discussed. Additionally, an attempt will be made to extract the pseudoscalar glueball mass from the TCDC at the matching Wilson flow time. This matching Wilson flow time is defined as the Wilson flow time that yields the best agreement between the topological charge density defined through gluonic and fermionic definitions. It is not intended to serve as a universally applicable or continuum-preferred choice, but rather as a practical benchmark to quantify the optimal matching of the Wilson flow when computing the gluonic definition of the topological charge density. We expect this benchmark to ensure computational requirements are met while conserving computational resources, particularly in large-scale calculations of topological charge density gluonic definition by using Wilson flow. This proof-of-concept study aims to assess the feasibility of this approach in lattice QCD and its potential applications in future large-scale statistical studies.

\section{Simulation setup}

The Lüscher-Weisz gauge action is used to generate the pure gauge lattice configurations. This gauge action is tadpole-improved at tree-level $\mathcal{O}\left(a^{2}\right)$ and combines the plaquette and rectangle gauge actions, implemented using the pseudo-heat-bath algorithm \cite{Luscher:1984xn,Bonnet:2001rc}. The parameters of the ensembles used to generate configurations with periodic boundary conditions are detailed in Tab. \ref{SPCU}, and the lattice spacings $a$ are determined through the Wilson flow with  $\omega_{0}=0.1670\left(10\right)\thinspace\text{fm}$ \cite{Luscher:2010iy,Sommer:2014mea,Borsanyi:2012zs}.

\begin{table}[H]
			\begin{centering}
				\begin{tabular*}{10cm}{@{\extracolsep{\fill}}ccccc}
					\hline 
					$\beta$ & $L^{3}\times T$ & $N_{\text{conf}}$ & $\omega\left[a\right]$ & $a\left(\text{fm}\right)$\tabularnewline
					\hline 
					$4.5$ & $16^{4}$ & $2000$ & $1.2978\left(40\right)$ & $0.1287\left(4\right)$\tabularnewline
					$4.8$ & $24^{3}\times48$ & $2000$ & $1.8648\left(28\right)$ & $0.0896\left(1\right)$\tabularnewline
					$5.0$ & $32^{4}$ & $1500$ & $2.5318\left(61\right)$ & $0.0660\left(2\right)$\tabularnewline
					\hline 
				\end{tabular*}
				\par\end{centering}
	\centering{}\caption{ Simulation parameters of configurations used. $N_{\text{conf}}$ is the total number of configurations in the ensemble. }
	\label{SPCU}
\end{table}

In this paper, we only use the overlap fermionic operator for the topological charge definition. The massless overlap Dirac operator ${D}_{\text{ov}}$ is given by
\begin{align}
	D_{\text{ov}} & =\left(\mathbbm{1}+\frac{D_{\rm{W}}}{\sqrt{D_{\rm{W}}^{\dagger}D_{\rm{W}}}}\right),
\end{align}
where $D_{\rm{W}}$ is the Wilson Dirac operator,
\begin{equation}
	D_{{\rm W}}  =\delta_{a,b}\delta_{\alpha,\beta}\delta_{i,j}-\kappa\sum_{\mu=1}^{4}[\left(\mathbbm{1}-\gamma_{\mu}\right)_{\alpha\beta} U_{\mu}\left(i\right)_{ab}\delta_{i,j-\hat{\mu}}+\left(\mathbbm{1}+\gamma_{\mu}\right)_{\alpha\beta}U_{\mu}^{\dagger}\left(i-\hat{\mu}\right)_{ab}\delta_{i,j+\hat{\mu}}],
\end{equation}
and $\kappa$ is the hopping parameter,
\begin{equation}
	\kappa=\frac{1}{2\left(-m+4\right)},
\end{equation}
which serves as an input parameter in the simulation. Since the results for multiple $\kappa$ values were discussed in detail in previous article \cite{Cheng:2020bql}, in this work, we will choose $\kappa=0.18$ and $0.19$ as representatives. While the topological charge computed using this overlap operator with the point sources yields an integer value, the computational cost is significantly high. To reduce the computational cost, the SMP method is introduced to calculate the topological charge density with the Dirac operator. The SMP method is utilized to calculate the topological charge density of the fermionic definition, as follows  \cite{Xiong:2019pmh,Cheng:2020bql}
\begin{align}
	q_{\text{smp}}\left(x\right) & =\sum_{\alpha,a}\psi\left(x,\alpha,a\right)\left(\tilde{D}_{\text{ov}}\left(x\right)\right)\phi_{P}\left(S\left(x,P\right),\alpha,a\right)\nonumber \\
	& =\sum_{\alpha,a}\psi\left(x,\alpha,a\right)\left(\tilde{D}_{\text{ov}}\left(x\right)\right)\psi\left(x,\alpha,a\right)\nonumber \\
	& \quad+\sum_{y\in S\left(x,P\right)}^{y\ne x}\psi\left(x,\alpha,a\right)\left(\tilde{D}_{\text{ov}}\left(x\right)\right)\psi\left(x,\alpha,a\right), \label{q-smp}
\end{align}
where $\tilde{D}_{\text{ov}}$ is defined as ${{\widetilde{D}}_{\text{ov}}}=\frac{1}{2}{{\gamma }_{5}}{{D}_{\text{ov}}}$. The SMP source vector $\phi_{P}\left(S\left(X,P\right),\alpha,a\right)$ is introduced as follows
\begin{equation}
	\phi_{P}\left(S\left(X,P\right),\alpha,a\right)=\mathop{\underset{y\in S\left(x,P\right)}{\sum}\psi\left(y,\alpha,a\right)},
\end{equation}
where $x$ is the seed site at $\left(x_{1},x_{2},x_{3},x_{4}\right)$ and $y$ represents the other lattice sites belongs to the set $S\left(x,P\right)$. $S\left(x,P\right)$ is the site with the same color of $x$ obtained by the symmetric coloring scheme $P\left(\frac{n_{s}}{d},\frac{n_{s}}{d},\frac{n_{s}}{d},\frac{n_{t}}{d},\text{mode}\right)$. $n_s$ and $n_d$ are the spatial and temporal sizes of lattice, $d$ is the minimal distance of the coloring scheme, $\text{mode = 0,1,2}$ corresponds to the Normal, Split and Combined mode for scheme $P$. The total number of SMP source vectors $N_\text{SMPV}$ that cover all lattice sites for the $(d,\text{mode})$ is
\[
N_{\text{SMPV}}=\begin{cases}
	d^{4}, & \text{mode}  = 0,\\
	2d^{4}, & \text{mode} = 1,\\
	\frac{d^{4}}{2}, & \text{mode} = 2.
\end{cases}
\]
Because of the space-time locality of ${D}_{\text{ov}}(x)$, the last line in Eq. (\ref{q-smp}) can be neglected. The topological charge density using the SMP source vector with the proper $P$ is 
\begin{equation}
	q_{\text{smp}}\left(x\right)=\sum_{\alpha,a}\psi\left(x,\alpha,a\right)\left(\tilde{D}_{\text{ov}}\left(x\right)\right)\psi\left(x,\alpha,a\right),
\end{equation}
and the topological charge for the SMP source vector is
\begin{equation}
Q_{\text{smp}}=\sum_{x}q_{\text{smp}}\left(x\right),
\end{equation}

The field tensor used in calculating the topological charge density of the gluonic definition is a 3-loop $\mathcal{O}\left(a^{4}\right)$-improved and defined as \cite{BilsonThompson:2002jk}, 
\begin{equation}
F_{\mu\nu}^{\text{Imp}}=\frac{27}{18}C_{\text{\ensuremath{\mu\nu}}}^{\left(1,1\right)}-\frac{27}{180}C_{\text{\ensuremath{\mu\nu}}}^{\left(2,2\right)}+\frac{1}{90}C_{\text{\ensuremath{\mu\nu}}}^{\left(3,3\right)},
\end{equation}
and $C_{\mu \nu }^{\left( m,m \right)}$ is the clover term constructed by $m \times m$ loops. The Wilson flow is used to smooth the gauge field in the gluonic definition, and the gauge fields do not need to be renormalized. The topological charge density of the gluonic definition by using the Wilson flow is 

\begin{equation}
{{q}_{\text{wf}}}\left( x \right)=\frac{1}{32{{\pi }^{2}}}{{\varepsilon }_{\mu \nu \rho \sigma }}\text{Tr}\left[ F_{\mu \nu }^{\text{Imp}}\left( x \right)F_{\rho \sigma }^{\text{Imp}}\left( x \right) \right],
\end{equation}
and the corresponding topological charge $Q_\text{wf}$ for the gluonic definition is given by
\begin{equation}
Q_\text{wf}=\sum\limits_{x}{{{q}_{\text{wf}}}}\left( x \right). \label{Q_wf}
\end{equation}

The matching Wilson flow time $\tau_{\text{qmr}}$ for calculating the topological charge using the gluonic definition is determined by minimizing the absolute difference between the topological charge obtained from the gluonic definition at various Wilson flow times and that calculated from the fermionic definition. Such a comparison demonstrates the global similarity of topological charge density. This process involves finding the minimum value of the following expression for different Wilson flow times,

\begin{equation}
\text{min}\left|Q-Q_{\text{wf}}\right|, \label{eq:minsub}
\end{equation}
where $Q$ is the topological charge obtained by rounding the topological charge $Q_\text{smp}$ calculated using the SMP method to the nearest integer, and $Q_{\text{wf}}$ represents the topological charge of the gluonic definition by using  the Wilson flow. When the above equation is satisfied, the corresponding Wilson flow time is the matching Wilson flow time $\tau_\text{qmr}$ for computing the topological charge of the gluonic definition. 

To investigate how to use the SMP method to determine the matching Wilson flow time $\tau_{\text{mr}}$ when computing the TCDC of the gluonic definition, a matching procedure is introduced. The matching paramter $\Xi_{AB}$, used to characterize the variation patterns of topological charge density at local positions calculated by two methods, will be calculated as follows \cite{Bruckmann:2006wf,Moran:2010rn},

\begin{equation}
\Xi_{AB}=\frac{\chi_{AB}^{2}}{\chi_{AA}\chi_{BB}},
\end{equation}
with
\begin{equation}
\chi_{AB}=\frac{1}{V}\sum_{x}\left(q_{A}\left(x\right)-\bar{q}_{A}\right)\left(q_{B}\left(x\right)-\bar{q}_{B}\right),
\end{equation}
where $\bar{q}$ is the mean value of topological charge density $q\left(x\right)$, and $V$ is the volume. Clearly, if ${{q}_{A}}\left( x \right)={{q}_{B}}\left( x \right)$, the matching parameter is $1$. When the numerical value of $\Xi_{AB}$ is nearest to $1$, which indicates that the local variation patterns of topological charge density exhibit the highest degree of consistency, and the flow time is the desired matching Wilson flow time $\tau_{\text{mr}}$ to calculate the TCDC of the gluonic definition.

\section{The matching parameter and the matching flow time of the Wilson flow \label{sec:The-matching-parameters} }
Before discussing the matching Wilson flow time for TCDC, we first examine the limitations of the matching Wilson flow time in calculating the topological charge of the gluonic definition. We also discuss the feasibility and advantages of the SMP method in computing the topological charge of the fermionic definition.

Although we can reduce the computational resources required for calculating the topological charge (density) of the fermion definition (overlap operator) by using the SMP method, when the SMP calculation parameters are $(8,0)$ and when calculating multiple groups $(d,\text{mode})$, the computational resources remain a significant challenge. Due to computational resource limitations, 10 configurations randomly selected from the first 100 configurations were analyzed for the $16^4$ and $24^3\times 48$ lattice ensembles at $\kappa = 0.18$ and $0.19$, while only 5 configurations were analyzed for $32^4$ lattice at $\kappa = 0.18$. The topological charge calculated using the fermionic definition by the SMP method at $\kappa = 0.18$ and the gluonic definition at the matching Wilson flow time $\tau_{\text{mr}}$ (determined by the mathching procedure) for 5 configurations of each lattice are presented as representatives in Fig. \ref{Q-diff-1}, and results for the remaining data are provided in Fig. \ref{Q-diff-2} of Appendix \ref{app-A}. It indicates that the SMP method with parameters $(8,0)$ achieves extremely high accuracy in calculating the topological charge defined by the fermionic method, which can be considered equivalent to the results obtained using point sources as the source vectors in the SMP method. Notably, when the parameters of the SMP source vectors are set to $(4,2)$, the topological charge $Q$ derived from the fermionic definition using the SMP method closely matches that calculated from the parameters $(8,0)$. 

\begin{figure}[H]
	\noindent \begin{centering}
		\includegraphics[scale=0.64]{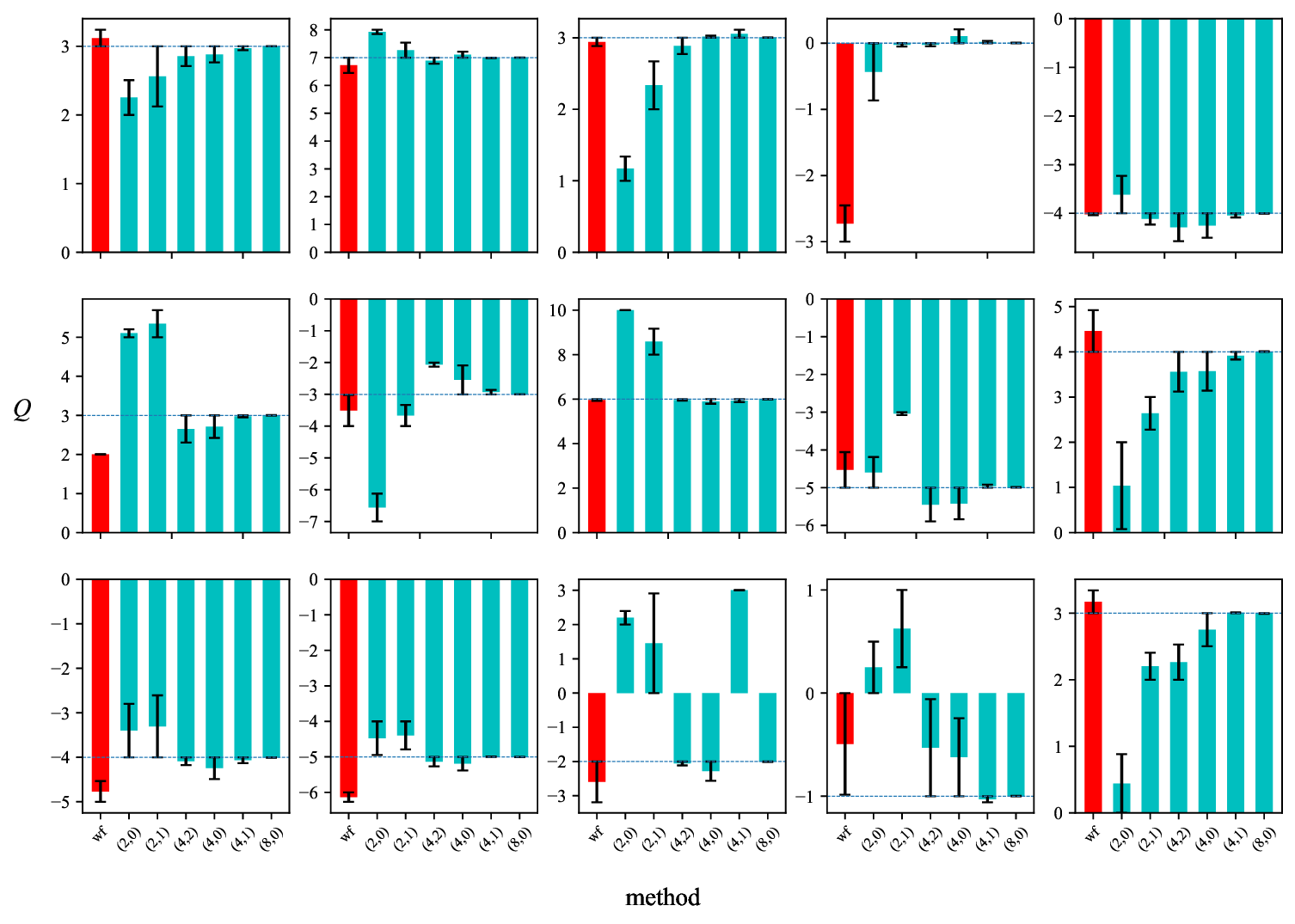}
		\par\end{centering}
	\caption{ The topological charge of the configurations calculated using the fermionic definition by the SMP method and the gluonic definition at the matching Wilson flow time $\tau_{\text{mr}}$. wf stands for the Wilson flow at $\tau_{\text{mr}}$. The error bars do not represent the uncertainties. Instead, they indicate the deviations of the topological charges $Q_{\text{smp}}$ calculated using the SMP method with different parameters $(d,\text{mode})$ and $Q_{\text{wf}}$ using the Wilson flow at the matching Wilson flow time from the topological charge $Q$. The integer topological charges $Q$ (the dashed line in the figure) are obtained by rounding the fermionic topological charges computed using the SMP method with parameters (8,0). The results for the lattices $16^{4}$, $24^{3}\times48$, and $32^{4}$ are presented from top to bottom, and the configurations for each lattice are shown from left to right.}
	\label{Q-diff-1}
\end{figure}

These findings indicate that using SMP sources with parameters $(4,2)$ could be a viable approach for calculating the topological charge of the fermionic definition, potentially reducing computational resource requirements compared to traditional point source calculations. All results suggest that the SMP method with parameters $(4,2)$ could be considered for application to the computation of topological charge of fermionic definition. In calculating the topological charge of the gluonic definition using the Wilson flow, the matching Wilson flow time can be determined using the Eq. (\ref{eq:minsub}). However, the topological charges of the gluonic definition, calculated at the matching Wilson flow time $\tau_{\text{mr}}$ through the matching procedure, deviate considerably from integer values. The upcoming research will demonstrate that the matching Wilson flow time determined by the matching method is not suitable for the calculation of topological charge, it is applicable for the calculation of TCDC. 

The topological charges of the fermionic definition calculated by the SMP method with $\kappa = 0.18$, the topological charges calculated using the gluonic definition by the Wilson flow, and the matching Wilson flow time $\tau_{\text{qmr}}$ are presented in Tab. \ref{Q-tao-pr-1}. The results of the same configrations for $\kappa = 0.19$ are given in the Tab. \ref{Q-tao-pr-3} of Appendix  \ref{app-A}. The value of $Q$ is obtained by rounding the topological charge $Q_{\text{smp}}$, calculated using the SMP method with parameters $(8,0)$, to the nearest integer. $Q_{\text{wf}}$ represents the topological charge obtained from the gluonic definition using the Wilson flow, rounded to the nearest integer. The matching Wilson flow time $\tau_{\text{qmr}}$ for the topological charge of gluonic definition is determined by identifying the minimum value of the absolute difference between $Q$ and $Q_{\text{wf}}$ as outlined in Eq. (\ref{eq:minsub}). The results demonstrate that the topological charge derived from the SMP method with parameters $(8,0)$ is numerically close to the precise value obtained from point sources. This indicates that $Q$ accurately represents the topological charge of the configuration.

The topological charge calculated using the SMP method $(\kappa = 0.18)$ with parameters $(4,2)$ and the Wilson flow, along with the matching Wilson flow time $\tau_{\text{qmr}}$ for the topological charge of the gluonic definition, is presentedin Tab. \ref{Q-tao-pr-2}, and the results for the same configurations $(\kappa = 0.19)$ are shown in the Tab. \ref{Q-tao-pr-4} of the Appendix \ref{app-A}. The methods for determining $Q$ and $\tau_{\text{qmr}}$ are the same as those employed in Tab. \ref{Q-tao-pr-1}. The values of $Q$ and $\tau_{\text{qmr}}$ obtained using the SMP method with parameters $(4,2)$ are fundamentally similar to those derived with parameters $(8,0)$. 

\begin{table}[H]
	\begin{tabular}{|m{1.9cm}|m{1.9cm}|m{1.9cm}|m{1.9cm}|m{1.9cm}|m{1.9cm}|m{1.9cm}|m{1.9cm}|}
		\hline 
		\makecell[c]{$L^3\times T$} & \makecell[c]{S/N} & \makecell[c]{$Q$} & \makecell[c]{$Q_{\text{smp}}$} & \makecell[c]{$\left|Q-Q_{\text{smp}}\right|$} & \makecell[c]{$Q_{\text{wf}}$} & \makecell[c]{$\left|Q-Q_{\text{wf}}\right|$} & \makecell[c]{$\tau_{\text{qmr}}$} \tabularnewline
		\hline 
		\makecell[c]{\multirow{5}{*}{$16^{4}$}} & \makecell[c]{$10$} & \makecell[c]{$3$} & \makecell[c]{$3.0002$} & \makecell[c]{$0.0002$} & \makecell[c]{$2.9823$} & \makecell[c]{$0.0177$} & \makecell[c]{$0.40$} \tabularnewline
		\cline{2-8} \cline{3-8} \cline{4-8} \cline{5-8} \cline{6-8} \cline{7-8} \cline{8-8} 
		& \makecell[c]{$18$} & \makecell[c]{$7$} & \makecell[c]{$7.0016$} & \makecell[c]{$0.0016$}  & \makecell[c]{$6.8060$} & \makecell[c]{$0.1942$} & \makecell[c]{$1.00$} \tabularnewline
		\cline{2-8} \cline{3-8} \cline{4-8} \cline{5-8} \cline{6-8} \cline{7-8} \cline{8-8} 
		& \makecell[c]{$20$} & \makecell[c]{$3$} & \makecell[c]{$3.0019$} & \makecell[c]{$0.0019$}  & \makecell[c]{$3.0365$} & \makecell[c]{$0.0365$} & \makecell[c]{$0.40$} \tabularnewline
		\cline{2-8} \cline{3-8} \cline{4-8} \cline{5-8} \cline{6-8} \cline{7-8} \cline{8-8} 
		& \makecell[c]{$22$} & \makecell[c]{$0$} & \makecell[c]{$0.0037$} & \makecell[c]{$0.0037$}  & \makecell[c]{$-0.8162$} & \makecell[c]{$0.8162$} & \makecell[c]{$1.00$} \tabularnewline
		\cline{2-8} \cline{3-8} \cline{4-8} \cline{5-8} \cline{6-8} \cline{7-8} \cline{8-8}
		& \makecell[c]{$30$} & \makecell[c]{$-4$} & \makecell[c]{$-4.0045$} & \makecell[c]{$0.0045$}  & \makecell[c]{$-4.0206$}  & \makecell[c]{$0.0206$} & \makecell[c]{$0.38$} \tabularnewline
		\hline 
		\makecell[c]{\multirow{5}{*}{$24^{3}\times48$}} & \makecell[c]{$5$} & \makecell[c]{$3$} & \makecell[c]{$3.0028$} & \makecell[c]{$0.0028$} & \makecell[c]{$2.9521$}  & \makecell[c]{$0.0479$}  & \makecell[c]{$1.00$} \tabularnewline
		\cline{2-8} \cline{3-8} \cline{4-8} \cline{5-8} \cline{6-8} \cline{7-8} \cline{8-8} 
		& \makecell[c]{$15$} & \makecell[c]{$-3$} & \makecell[c]{$3.0021$} & \makecell[c]{$0.0021$}  & \makecell[c]{$-2.9871$}  & \makecell[c]{$0.0129$}  & \makecell[c]{$0.26$} \tabularnewline
		\cline{2-8} \cline{3-8} \cline{4-8} \cline{5-8} \cline{6-8} \cline{7-8} \cline{8-8} 
		& \makecell[c]{$17$} & \makecell[c]{$6$} & \makecell[c]{$5.9946$} & \makecell[c]{$0.0054$}  & \makecell[c]{$5.9620$}  & \makecell[c]{$0.0380$}  & \makecell[c]{$0.34$} \tabularnewline
		\cline{2-8} \cline{3-8} \cline{4-8} \cline{5-8} \cline{6-8} \cline{7-8} \cline{8-8}
		& \makecell[c]{$30$} & \makecell[c]{$-5$} & \makecell[c]{$-4.9933$}  & \makecell[c]{$0.0067$}  & \makecell[c]{$-4.9756$} & \makecell[c]{$0.0244$}  & \makecell[c]{$0.44$} \tabularnewline
		\cline{2-8} \cline{3-8} \cline{4-8} \cline{5-8} \cline{6-8} \cline{7-8} \cline{8-8}
		& \makecell[c]{$53$} & \makecell[c]{$4$} & \makecell[c]{$4.0057$}  & \makecell[c]{$0.0057$}  & \makecell[c]{$4.0062$} & \makecell[c]{$0.0062$}  & \makecell[c]{$0.26$} \tabularnewline
		\hline 
		\makecell[c]{\multirow{5}{*}{$32^{4}$}} & \makecell[c]{$22$} & \makecell[c]{$-4$} & \makecell[c]{$-4.0043$} & \makecell[c]{$0.0043$} & \makecell[c]{$-4.0174$} & \makecell[c]{$0.0174$} & \makecell[c]{$1.00$} \tabularnewline
		\cline{2-8} \cline{3-8} \cline{4-8} \cline{5-8} \cline{6-8} \cline{7-8} \cline{8-8} 
		& \makecell[c]{$25$} & \makecell[c]{$-5$} & \makecell[c]{$-5.0033$} & \makecell[c]{$0.0033$} & \makecell[c]{$-5.0165$} & \makecell[c]{$0.0165$} & \makecell[c]{$1.00$} \tabularnewline
		\cline{2-8} \cline{3-8} \cline{4-8} \cline{5-8} \cline{6-8} \cline{7-8} \cline{8-8}
		& \makecell[c]{$28$} & \makecell[c]{$-2$} & \makecell[c]{$-2.0057$} & \makecell[c]{$0.0057$}  & \makecell[c]{$-1.9953$}  & \makecell[c]{$0.0047$}  & \makecell[c]{$0.14$} \tabularnewline
		\cline{2-8} \cline{3-8} \cline{4-8} \cline{5-8} \cline{6-8} \cline{7-8} \cline{8-8}
		& \makecell[c]{$40$} & \makecell[c]{$-1$} & \makecell[c]{$-0.9987$} & \makecell[c]{$0.0013$}  & \makecell[c]{$-0.9974$}  & \makecell[c]{$0.0026$}  & \makecell[c]{$0.24$} \tabularnewline
		\cline{2-8} \cline{3-8} \cline{4-8} \cline{5-8} \cline{6-8} \cline{7-8} \cline{8-8} 
		& \makecell[c]{$50$} & \makecell[c]{$3$} & \makecell[c]{$2.9963$} & \makecell[c]{$0.0037$} & \makecell[c]{$3.0008$} & \makecell[c]{$0.0008$} & \makecell[c]{$1.00$} \tabularnewline
		\hline 
	\end{tabular}
	
	\caption{ The matching Wilson flow time $\tau_{\text{qmr}}$ for the topological charge $Q_{\text{wf}}$ calculated using the Wilson flow. $Q$ is obtained by rounding the result $Q_{\text{smp}}$ of the SMP method with parameters $(8,0)$ to the nearest integer and $\kappa = 0.18$. S/N is the sequence number of the configuration.}
	
	\label{Q-tao-pr-1}
\end{table}

\begin{table}[H]
	\begin{tabular}{|m{1.9cm}|m{1.9cm}|m{1.9cm}|m{1.9cm}|m{1.9cm}|m{1.9cm}|m{1.9cm}|m{1.9cm}|}
		\hline 
		\makecell[c]{$L^3\times T$} & \makecell[c]{S/N} & \makecell[c]{$Q$} & \makecell[c]{$Q_{\text{smp}}$} & \makecell[c]{$\left|Q-Q_{\text{smp}}\right|$} & \makecell[c]{$Q_{\text{wf}}$} & \makecell[c]{$\left|Q-Q_{\text{wf}}\right|$} & \makecell[c]{$\tau_{\text{qmr}}$} \tabularnewline
		\hline 
		\makecell[c]{\multirow{5}{*}{$16^{4}$}} & \makecell[c]{$10$} & \makecell[c]{$3$} & \makecell[c]{$2.8563$} & \makecell[c]{$0.1437$} & \makecell[c]{$2.9823$} & \makecell[c]{$0.0177$} & \makecell[c]{$0.40$} \tabularnewline
		\cline{2-8} \cline{3-8} \cline{4-8} \cline{5-8} \cline{6-8} \cline{7-8} \cline{8-8} 
		& \makecell[c]{$18$} & \makecell[c]{$7$} & \makecell[c]{$6.8914$} & \makecell[c]{$0.1086$} & \makecell[c]{$6.8058$} & \makecell[c]{$0.1942$} & \makecell[c]{$1.00$} \tabularnewline
		\cline{2-8} \cline{3-8} \cline{4-8} \cline{5-8} \cline{6-8} \cline{7-8} \cline{8-8} 
		& \makecell[c]{$20$} & \makecell[c]{$3$} & \makecell[c]{$2.8864$} & \makecell[c]{$0.1136$} & \makecell[c]{$3.0365$} & \makecell[c]{$0.0365$} & \makecell[c]{$0.40$} \tabularnewline
		\cline{2-8} \cline{3-8} \cline{4-8} \cline{5-8} \cline{6-8} \cline{7-8} \cline{8-8} 
		& \makecell[c]{$22$} & \makecell[c]{$0$} & \makecell[c]{$-0.0238$} & \makecell[c]{$0.0238$} & \makecell[c]{$-0.8162$} & \makecell[c]{$0.8162$} & \makecell[c]{$1.00$} \tabularnewline
		\cline{2-8} \cline{3-8} \cline{4-8} \cline{5-8} \cline{6-8} \cline{7-8} \cline{8-8} 
		& \makecell[c]{$30$} & \makecell[c]{$-4$} & \makecell[c]{$-4.2867$} & \makecell[c]{$0.2867$} & \makecell[c]{$-4.0206$} & \makecell[c]{$0.0206$} & \makecell[c]{$0.38$} \tabularnewline
		\hline 
		\makecell[c]{\multirow{5}{*}{$24^{3}\times48$}} & \makecell[c]{$5$} & \makecell[c]{$3$} & \makecell[c]{$2.6528$} & \makecell[c]{$0.3472$} & \makecell[c]{$2.9521$} & \makecell[c]{$0.0479$} & \makecell[c]{$1.00$} \tabularnewline
		\cline{2-8} \cline{3-8} \cline{4-8} \cline{5-8} \cline{6-8} \cline{7-8} \cline{8-8} 
		& \makecell[c]{$15$} & \makecell[c]{$-2$} & \makecell[c]{$-2.0643$} & \makecell[c]{$0.9357$} & \makecell[c]{$-2.1213$} & \makecell[c]{$0.8787$} & \makecell[c]{$0.18$} \tabularnewline
		\cline{2-8} \cline{3-8} \cline{4-8} \cline{5-8} \cline{6-8} \cline{7-8} \cline{8-8} 
		& \makecell[c]{$17$} & \makecell[c]{$6$} & \makecell[c]{$5.9656$} & \makecell[c]{$0.0344$} & \makecell[c]{$5.9620$} & \makecell[c]{$0.0380$} & \makecell[c]{$0.34$} \tabularnewline
		\cline{2-8} \cline{3-8} \cline{4-8} \cline{5-8} \cline{6-8} \cline{7-8} \cline{8-8}
		& \makecell[c]{$30$} & \makecell[c]{$-5$} & \makecell[c]{$-5.4482$} & \makecell[c]{$0.4482$} & \makecell[c]{$-4.9756$} & \makecell[c]{$0.0244$} & \makecell[c]{$0.44$}\tabularnewline
		\cline{2-8} \cline{3-8} \cline{4-8} \cline{5-8} \cline{6-8} \cline{7-8} \cline{8-8}
		& \makecell[c]{$53$} & \makecell[c]{$4$} & \makecell[c]{$3.5592$} & \makecell[c]{$0.4408$} & \makecell[c]{$4.0062$} & \makecell[c]{$0.0062$} & \makecell[c]{$0.26$}\tabularnewline
		\hline 
		\makecell[c]{\multirow{5}{*}{$32^{4}$}} & \makecell[c]{$22$} & \makecell[c]{$-4$} & \makecell[c]{$-4.0877$} & \makecell[c]{$0.0877$} & \makecell[c]{$-4.0174$} & \makecell[c]{$0.0174$} & \makecell[c]{$1.00$}\tabularnewline
		\cline{2-8} \cline{3-8} \cline{4-8} \cline{5-8} \cline{6-8} \cline{7-8} \cline{8-8} 
		& \makecell[c]{$25$} & \makecell[c]{$-5$} & \makecell[c]{$-5.1343$} & \makecell[c]{$0.1343$} & \makecell[c]{$-5.0165$} & \makecell[c]{$0.0165$} & \makecell[c]{$1.00$}\tabularnewline
		\cline{2-8} \cline{3-8} \cline{4-8} \cline{5-8} \cline{6-8} \cline{7-8} \cline{8-8} 
		& \makecell[c]{$28$} & \makecell[c]{$-2$} & \makecell[c]{$-2.0568$} & \makecell[c]{$0.0568$} & \makecell[c]{$-1.9953$} & \makecell[c]{$0.0047$} & \makecell[c]{$0.14$}\tabularnewline
		\cline{2-8} \cline{3-8} \cline{4-8} \cline{5-8} \cline{6-8} \cline{7-8} \cline{8-8} 
		& \makecell[c]{$40$} & \makecell[c]{$-1$} & \makecell[c]{$-0.5291$} & \makecell[c]{$0.4709$} & \makecell[c]{$-0.9974$} & \makecell[c]{$0.0026$} & \makecell[c]{$0.24$} \tabularnewline
		\cline{2-8} \cline{3-8} \cline{4-8} \cline{5-8} \cline{6-8} \cline{7-8} \cline{8-8} 
		& \makecell[c]{$50$} & \makecell[c]{$2$} & \makecell[c]{$2.2644$} & \makecell[c]{$0.7356$} & \makecell[c]{$3.0008$} & \makecell[c]{$0.0008$} & \makecell[c]{$1.00$} \tabularnewline
		\hline 
	\end{tabular}
	
	\caption{ The matching Wilson flow time $\tau_{\text{qmr}}$ for calculating $Q$ of the gluonic definition. $Q$ is obtained by rounding the result of the SMP method with parameters $(4,2)$ to the nearest integer and $\kappa = 0.18$. S/N stands for the sequence number of configurations.}
	
	\label{Q-tao-pr-2}
\end{table}

These results indicate that the SMP method with parameters $(4,2)$ is indeed a viable approach for accurately determining the topological charge of the fermionic definition. Moreover, this method can effectively establish the matching Wilson flow time $\tau_{\text{qmr}}$ when calculating the topological charge of the gluonic definition using the Wilson flow. Next, we will determine the matching Wilson flow time for calculating the TCDC of the gluonic definition through matching parameters.

In Ref. \cite{Cheng:2020bql}, the results indicate that the outcomes of the SMP method with parameters $(8,0)$ can serve as a benchmark for calculating the matching parameter $\Xi_{AB}$. To compare results from different parameters $(d\neq8,\text{mode}\neq0)$ with those obtained using the parameter set $(8,0)$ in the SMP method, the matching parameters $\Xi_{AB}$ are computed for various hopping parameters $\kappa$. 

\begin{table}[htbp!]
	\renewcommand{\arraystretch}{1.2} 
		\begin{centering}
			\begin{tabular}{|m{2.0cm}<{\centering}|m{2.2cm}<{\centering}|m{2.2cm}<{\centering}|m{2.2cm}<{\centering}|m{2.2cm}<{\centering}|m{2.2cm}<{\centering}|m{2.2cm}<{\centering}|}
				\hline 
				$\text{$L^3\times T$}$ & S/N & $\left(2,0\right)$ & $\left(2,1\right)$ & $\left(4,2\right)$ & $\left(4,0\right)$ & $\left(4,1\right)$\\ \hline
				
				\multirow{10}{*}{\centering $16^{4}$} 
				& \multirow{2}{*}{10} 
				& 0.638631 & 0.819229 & 0.976946 & 0.988357 & 0.999246 \\
				& & (11391) & (13795) & (15949) & (16119) & (16261) \\ \cline{2-7}
				& \multirow{2}{*}{18} 
				& 0.625724 & 0.810913 & 0.975660 & 0.987727 & 0.9991685 \\
				& & (10871) & (13406) & (15644) & (15798) & (15978) \\ \cline{2-7}
				& \multirow{2}{*}{20}  
				& 0.630302 & 0.814710 & 0.976150 & 0.987895 & 0.999242 \\
				& & (11290) & (13939) & (16344) & (16516) & (16636) \\ \cline{2-7}
				& \multirow{2}{*}{22}  
				& 0.629449 & 0.813132 & 0.976115 & 0.987805 & 0.999208 \\
				& & (10909) & (13406) & (15699) & (15868) & (16043) \\ \cline{2-7}
				& \multirow{2}{*}{30}  
				& 0.625374 & 0.812900 & 0.976215 & 0.988113 & 0.999250 \\ 
				& & (11037) & (13671) & (16037) & (16223) & (16371) \\ \hline
				
				\multirow{10}{*}{\centering $24^{3}\times48$} 
				& \multirow{2}{*}{5} 
				& 0.622905 & 0.811863 & 0.978178 & 0.988439 & 0.999436 \\
				& & (3237) & (3996) & (4671) & (4712) & (4757) \\ \cline{2-7}
				& \multirow{2}{*}{15} 
				& 0.621431 & 0.811156 & 0.978089 & 0.988359 & 0.999436 \\
				& & (3201) & (3954) & (4621) & (4664) & (4709) \\ \cline{2-7}
				& \multirow{2}{*}{17} 
				& 0.616351 & 0.807719 & 0.977245 & 0.988039 & 0.999396 \\
				& & (3229) & (4000) & (4689) & (4732) & (4778) \\ \cline{2-7}
				& \multirow{2}{*}{30} 
				& 0.622252 & 0.812087 & 0.978154 & 0.988425 & 0.999437 \\
				& & (3246) & (4013) & (4700) & (4743) & (4787) \\ \cline{2-7}
				& \multirow{2}{*}{53} 
				& 0.618310 & 0.808708 & 0.977281 & 0.988122 & 0.999397 \\
				& & (3247) & (4008) & (4701) & (4744) & (4789) \\ \hline
				
				\multirow{10}{*}{\centering $32^{4}$} 
				& \multirow{2}{*}{22} 
				& 0.619765 & 0.811184 & 0.978735 & 0.988528 & 0.999475 \\
				& & (2504) & (3102) & (3627) & (3658) & (3692) \\ \cline{2-7}
				& \multirow{2}{*}{25} 
				& 0.617173 & 0.810155 & 0.978567 & 0.988435 & 0.999472 \\
				& & (2486) & (3083) & (3609) & (3640) & (3675) \\ \cline{2-7}
				& \multirow{2}{*}{28} 
				& 0.617334 & 0.809961 & 0.978628 & 0.988456 & 0.999473 \\
				& & (2504) & (3099) & (3632) & (3664) & (3698) \\ \cline{2-7}
				& \multirow{2}{*}{40} 
				& 0.619947 & 0.811330 & 0.978676 & 0.988496 & 0.999475 \\
				& & (2497) & (3090) & (3616) & (3647) & (3681) \\ \cline{2-7}
				& \multirow{2}{*}{50} 
				& 0.618602 & 0.810218 & 0.978552 & 0.988459 & 0.999470 \\
				& & (2496) & (3090) & (3618) & (3649) & (3685) \\ \hline
			\end{tabular}
			\par\end{centering}
		\caption{$\Xi_{AB}$ for SMP method, obtained by comparing $(d\ne8,\text{mode}\ne0)$ to $(8,0)$. The hopping parameter
			is $\kappa=0.18$. S/N stands for the sequence number of configurations. The values in parentheses indicate the standard deviation. To distinguish the computed results, six significant digits were used.}
		\label{Tab-1}
\end{table}

The matching parameters of SMP method, determined by comparing parameters $(d \ne 8, \text{mode} \ne 0)$ to $(8,0)$, for five representative configurations at $\kappa=0.18$ on lattices $16^{4}$, $24^{3}\times48$, and $32^{4}$ are provided in Tab. \ref{Tab-1}. Tab. \ref{Tab-2} presents the matching parameters $\Xi_{AB}$ for the same five representative configurations of the $16^{4}$ and $24^{3}\times48$ lattices at $\kappa=0.19$. The matching parameter results $\Xi_{AB}$  for the remaining configurations are provided in the Tab. \ref{xi_AB_app_1} and \ref{xi_AB_app_2} of Appendix \ref{app-A}. It can be observed that as the SMP parameter $d$ (the number of source vectors) increases, the matching parameter also increases. This is expected because, with a larger $d$ (the number of source vectors), the terms neglected in Eq. (\ref{q-smp}) (the error) become smaller. When the SMP parameters are $(4,2)$, the matching parameters are already very close to $1$, indicating that the topological charge density calculated with parameters $(4,2)$ can also be used to determine the matching parameter $\Xi_{AB}$. Using the SMP method with parameters $(4,2)$ requires approximately $1/32$ of the computational resources compared to the parameters $(8,0)$, or $1/512$ of the computational resources compared to the point sources, leading to significant resource savings.

\begin{table}[htbp!]
		\renewcommand{\arraystretch}{1.2} 
		\begin{centering}
			\begin{tabular}{|m{2.0cm}<{\centering}|m{2.2cm}<{\centering}|m{2.2cm}<{\centering}|m{2.2cm}<{\centering}|m{2.2cm}<{\centering}|m{2.2cm}<{\centering}|m{2.2cm}<{\centering}|}
				\hline 
				$\text{$L^3\times T$}$ & S/N & $\left(2,0\right)$ & $\left(2,1\right)$ & $\left(4,2\right)$ & $\left(4,0\right)$ & $\left(4,1\right)$\\ \hline
				
				\multirow{10}{*}{\centering $16^{4}$} 
				& \multirow{2}{*}{10} 
				& 0.681956 & 0.853189 & 0.984066 & 0.991147 & 0.999579 \\
				& & (12443) & (14940) & (16819) & (16916) & (17040) \\ \cline{2-7}
				& \multirow{2}{*}{18} 
				& 0.668659 & 0.845414 & 0.983047 & 0.990692 & 0.999539 \\
				& & (11853) & (14347) & (16335) & (16446) & (16603) \\ \cline{2-7}
				& \multirow{2}{*}{20}  
				& 0.671241 & 0.847403 & 0.983229 & 0.990655 & 0.999558 \\
				& & (11979) & (14543) & (16602) & (16699) & (16796) \\ \cline{2-7}
				& \multirow{2}{*}{22}  
				& 0.671480 & 0.846203 & 0.983362 & 0.990746 & 0.999551 \\
				& & (11806) & (14292) & (16296) & (16417) & (16549) \\ \cline{2-7}
				& \multirow{2}{*}{30}  
				& 0.666107 & 0.844403 & 0.983037 & 0.990749 & 0.999552 \\ 
				& & (11731) & (14279) & (16333) & (16454) & (16572) \\ \hline
				
				\multirow{10}{*}{\centering $24^{3}\times48$} 
				& \multirow{2}{*}{5} 
				& 0.649081 & 0.833534 & 0.982238 & 0.990288 & 0.999565 \\
				& & (3356) & (4100)   & (4707)   &  (4739)  & (4777) \\ \cline{2-7}
				& \multirow{2}{*}{15} 
				& 0.647766 & 0.832886 & 0.982189 & 0.990227 & 0.999564 \\
				& & (3322) & (4063)   & (4665)   & (4698)   & (4737) \\ \cline{2-7}
				& \multirow{2}{*}{17} 
				& 0.644876  & 0.831116 & 0.981806 & 0.990044 & 0.999548 \\
				& & (3374) & (4137)   & (4757)   & (4790)   & (4830) \\ \cline{2-7}
				& \multirow{2}{*}{30} 
				& 0.648395 & 0.833703 & 0.982230 & 0.990281 & 0.999565 \\
				& & (3359) & (4113)  & (4731)    & (4765)   & (4804) \\ \cline{2-7}
				& \multirow{2}{*}{53} 
				& 0.646122 & 0.831612 & 0.981851 & 0.990116 & 0.999549 \\
				& & (3376) & (4123)  & (4749)    & (4782)   & (4821) \\ \hline
			\end{tabular}
\par\end{centering}

\caption{$\Xi_{AB}$ for SMP method, determined by comparing $(d\ne8,\text{mode}\ne0)$ to $(8,0)$ at $\kappa=0.19$. S/N represents the configuration number. The numbers in parentheses represent the standard deviation. Six significant digits are used to distinguish the computed results.}
\label{Tab-2}
\end{table}
		
In Fig. \ref{xi_diff-1} and Fig. \ref{xi_diff-2} of Appendix \ref{app-A}, the matching parameters and the matching Wilson flow time for the comparison between the topological charge density calculated using the SMP method (different parameters $(d,\text{mode})$) with the gluonic definition of the topological charge density at different Wilson flow times are presented. The results indicate that as the number of SMP source vectors increases, the value of the matching parameter $\text{\ensuremath{\Xi_{AB}}}$ also increases. This is expected, as increasing the number of SMP source vectors reduces the error in the SMP method. However, even when the number of SMP source vectors exceeds the source vector count at the parameters $(4,2)$, the increase in the matching parameters is no longer significant. This suggests that the SMP method with parameters $(4,2)$ can be used to determine the matching Wilson flow time for calculating the topological charge density in the gluonic definition. Additionally, the matching Wilson flow time remains essentially constant, regardless of variations in the parameters $(d,\text{mode})$ within the SMP method. The matching Wilson flow time in the calculation of topological density in the gluonic definition can be obtained from the matching procedure. All results indicate that choosing the parameters $(4,2)$ in the SMP method is a good option when selecting a benchmark to determine the matching parameter $\Xi_{AB}$.

\begin{figure}[htbp!]
\noindent \begin{centering}
\includegraphics[scale=0.54]{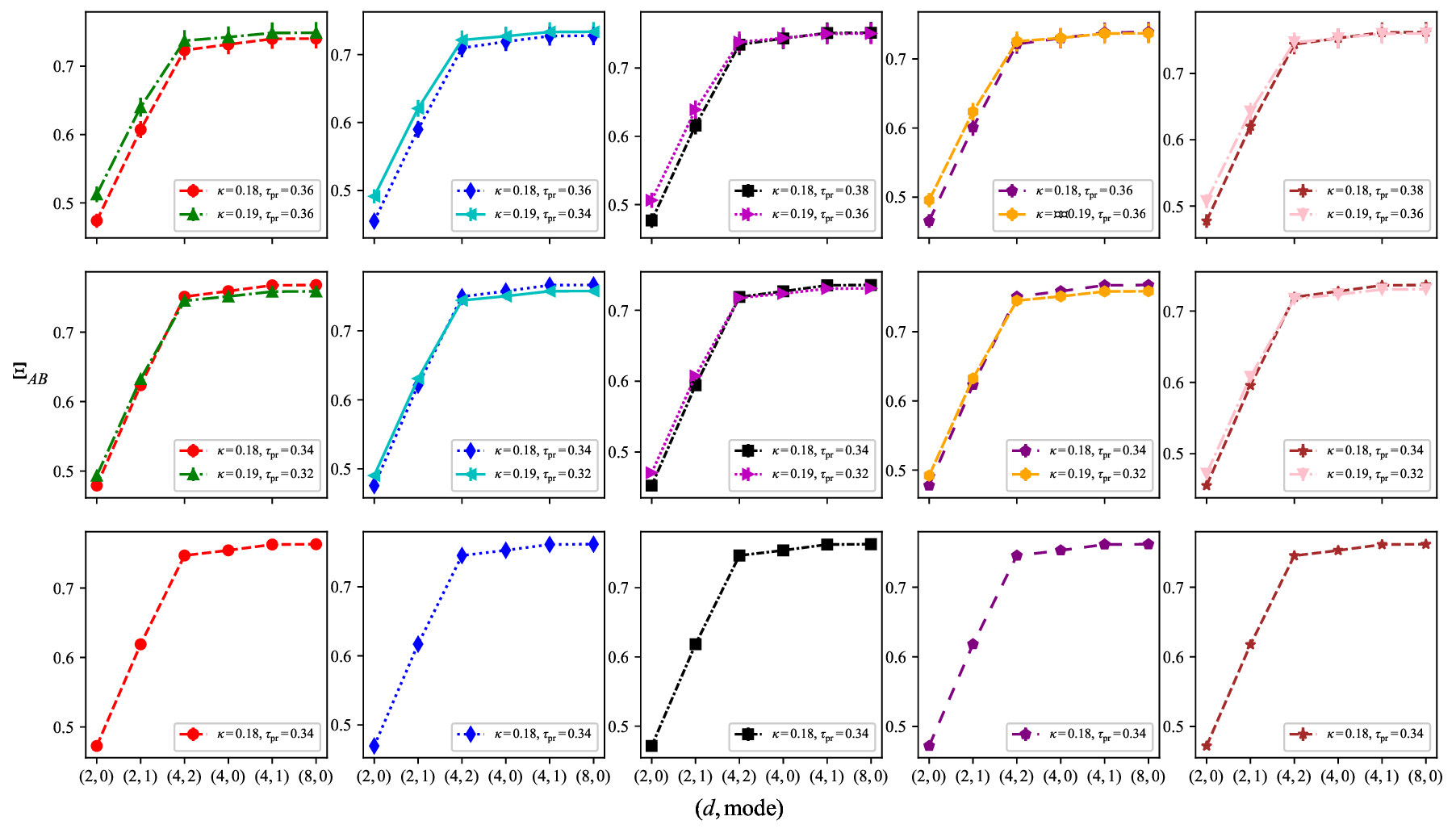}
\par\end{centering}
\caption{ The best $\text{\ensuremath{\Xi_{AB}}}$ for SMP with different parameters $\left(d,\text{mode}\right)$ compared with the Wilson flow in the calculation of topological charge density. $\tau_{\text{mr}}$ is the matching Wilson flow time, and $\kappa$ is the hopping parameter. From top to bottom, the results for the lattices $16^{4}$, $24^{3}\times48$, and $32^{4}$ are presented. However, due to computational resource limitations, results were only computed for $\kappa=0.18$ on the lattice $32^{4}$. The results of different configurations for each lattice are shown from left to right.}
\label{xi_diff-1}
\end{figure}

Furthermore, it suggests that a larger Wilson flow time is generally necessary when calculating the topological charge for the gluonic definition. This also indicates that the matching Wilson flow time $\tau_{\text{mr}}$ determined by the matching method may not be the optimal choice to calculate the topological charge of the gluonic definition. 

\begin{figure}
	\includegraphics[scale=0.60]{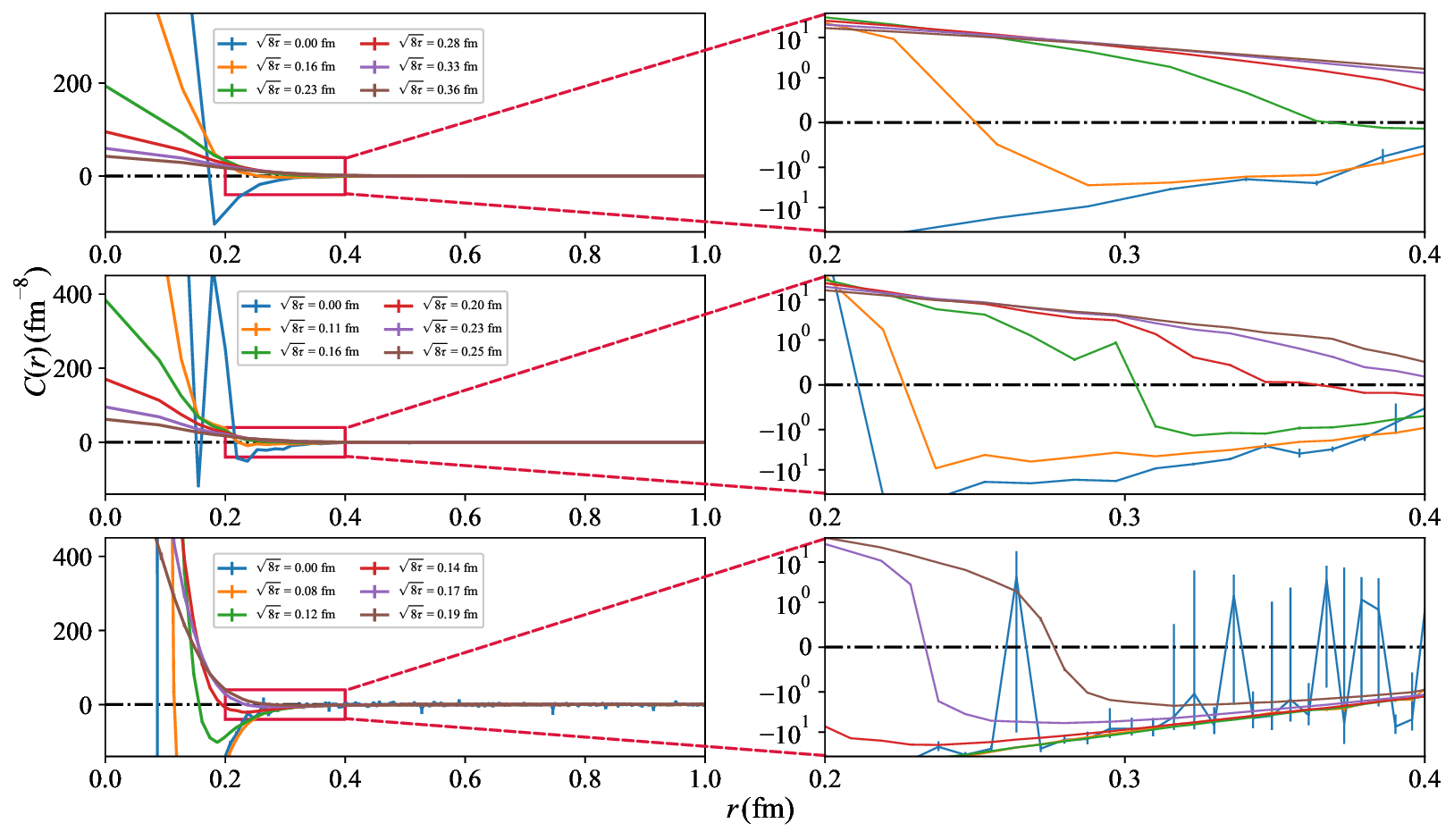}
	\caption{ $C\left(r/a\right)$ v.s. $r$ for different lattices. From top to bottom, the results for $16^{4}$ with $\beta=4.5$, $24^{3}\times48$ with $\beta=4.80$, and $32^{4}$ with $\beta=5.0$ are presented. $\sqrt{8\tau}$ is the flow radius of the Wilson flow time $\tau$. The right subplots are magnified views of the main plots on the left, focusing on the range \(0.2 \leq r \leq 0.4 \, \text{fm}\). The \texttt{symlog} scale in \texttt{matplotlib} is used for the y-axis, applying a linear scale in the range \([-1, 1]\) and a logarithmic scale outside this range.
	}
	\label{corr-diff}
\end{figure}

Notably, all results show that when calculating the topological charge of the gluonic definition, the matching Wilson flow time $\tau_{\text{qmr}}$ is generally larger than $\tau_{\text{mr}}$. The results demonstrate that when calculating TCDC at the matching Wilson flow time $\tau_{\text{qmr}}$, TCDC may exhibit oversmearing, meaning that the negative dip of TCDC may disappear, as shown in Fig. \ref{corr-diff}.

We adopt the matching results at $\kappa = 0.18$. In the calculation of the topological charge density, the matching Wilson flow time for lattices of $16^{4}$ at $\beta=4.5$, $24^{3}\times48$ at $\beta=4.8$ and $32^{4}$ at $\beta=5.0$ are approximately $\tau=0.38$, $\tau=0.34$ and $\tau=0.34$. The corresponding matching flow radii of the Wilson flow  $\sqrt{8\tau_\text{mr}}=0.224$, $0.148,$ and $0.109\thinspace\text{fm}$, respectively. In the subsequent calculation of TCDC, this matching Wilson flow time $\tau_{\text{mr}}$ will be used in the Wilson flow.

\section{The topological charge density correlator and the pseudoscalar glueball
mass\label{part:Topological-charge-density}}

TCDC is defined as \cite{Chowdhury:2012sq}

\begin{equation}
C\left(r\right)=\left\langle q\left(x\right)q\left(0\right)\right\rangle ,\thinspace\thinspace\thinspace\thinspace r=\left|x\right|,
\end{equation}
and the four-volume integral of the TCDC gives the topological susceptibility

\begin{equation}
\chi=\int\text{d}^{4}x\left\langle q\left(x\right)q\left(0\right)\right\rangle =\frac{\left\langle Q^{2}\right\rangle }{V},\thinspace\thinspace\thinspace\thinspace V\to\infty.
\end{equation}

When calculating the TCDC, the number of configurations for each lattice ensemble is given by $N_{\text{conf}}$ in Tab. \ref{SPCU}. Due to the presence of severe singularities and lattice artifacts in TCDC, a smoothing procedure is essential to refine the gauge fields. In this study, we employ the Wilson flow method for this purpose. Undersmearing fails to adequately eliminate the lattice artifacts, while oversmearing can erase even the negative character of the TCDC, as illustrated in Fig. \ref{corr-diff}. The results indicate that the flow times at which the negative dip disappears at a distance of $r\sim 0.3\,\text{fm}$ are $\tau \approx 0.4$ for $16^4$, $\tau \approx 0.6$ for $24^3\times48$, and $\tau \approx 1.0$ for $32^4$, which are very close to the corresponding $\tau_{\text{qmr}}$ determined by Eq. (\ref{eq:minsub}). In other words, when using the Wilson flow method to calculate TCDC at $\tau_{\text{qmr}}$, TCDC may be oversmearing.

The topological susceptibility $\chi$ for different lattice ensembles under the matching Wilson flow time $\tau_{\text{mr}}$ are presented in Tab. \ref{top_sus}, which are consistent with those in Ref. \cite{DelDebbio:2004ns,Luscher:2010ik,Athenodorou:2020ani}. This indicates that the matching Wilson flow time can also serve as a suitable choice for the flow time when calculating the topological susceptibility.

\begin{table}[H]
	\begin{spacing}{1.5}
	  \begin{centering}
	    \begin{tabular*}{10cm}{@{\extracolsep{\fill}}cccc}
		\hline 
		$L^{3}\times T$ & $16^{4}$ & $24^{3}\times48$ & $32^{4}$\tabularnewline
		\hline 
		$\chi\left[\text{MeV}^{4}\right]$ & $191\left(6\right)$ & $193\left(6\right)$ & $187\left(12\right)$\tabularnewline
		\hline 
	\end{tabular*}
    \par\end{centering}
    \end{spacing}
	\caption{The topological susceptibility $\chi$ of different lattice ensembles determined by the corresponding
		matching Wilson flow times $\tau_{\text{mr}}$. }
	\label{top_sus}
\end{table}

TCDC can be used to extract the lowest pseudoscalar glueball mass in the negative region by the following form

\begin{equation}
\left\langle q\left(x\right)q\left(0\right)\right\rangle =\frac{m}{4\pi^{2}r}K_{1}\left(mr\right),\label{eq:TCDC-Bess}
\end{equation}
and $K_{1}\left(z\right)$ is a modified Bessel function, which has the asymptotic form \cite{Shuryak:1994rr}

\begin{equation}
K_{1}\left(z\right)\underset{\text{large}\thinspace z}{\sim}e^{-z}\sqrt{\frac{\pi}{2z}}\left[1+\frac{3}{8z}\right].\label{eq:Bess-Lar-dis}
\end{equation}

\begin{figure}[htbp!]
\includegraphics[scale=0.52]{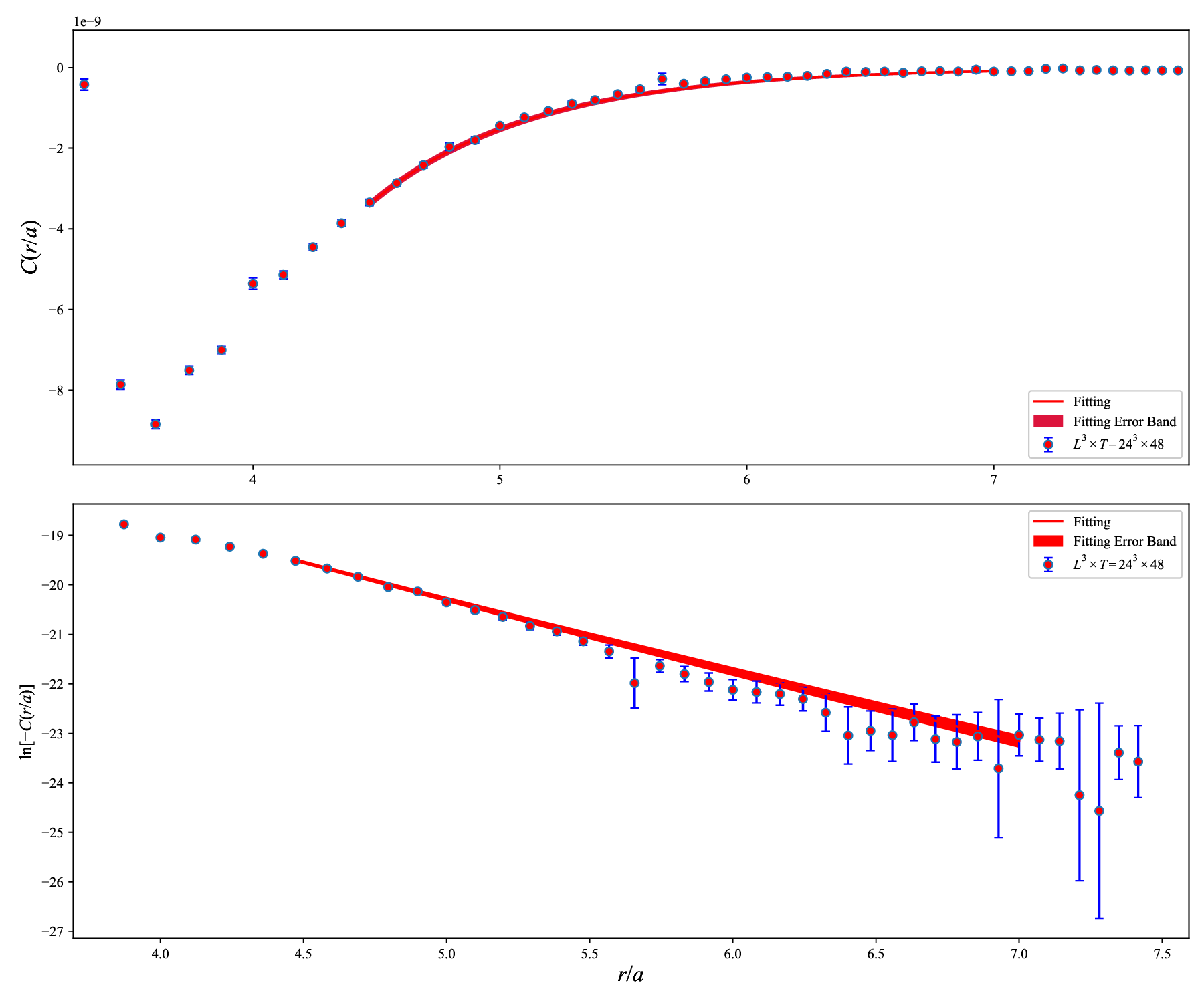}
\caption{ $C\left(r/a\right)$ and the local $\ln \left( -C\left( r \right) \right)$ versus $r$ at the matching Wilson flow time $\tau=0.34$ for the ensemble $24^{3}\times48$. The fitting curve and the fitting error to extract the pseudoscalar glueball mass is also shown.}

\label{C_r-vs-r}
\end{figure}

We aim to extract the glueball mass from the TCDC of three ensembles using the Wilson flow at the matching Wilson flow time $\tau_{\text{mr}}$. To do this, we apply Eq. (\ref{eq:TCDC-Bess}) and Eq. (\ref{eq:Bess-Lar-dis}) to determine the pseudoscalar glueball mass in the negative region by using a correlated $\chi^{2}$ fitting, considered the correlation of the data. In this fitting procedure, both the amplitude and mass are treated as free parameters, and the ${{{\chi }^{2}}}/{\text{dof}}$ is calculated to evaluate the quality of the fit. It shows that the extracted mass remains independent of the endpoint once the error bars of the tail of the TCDC approach zero \cite{Chowdhury:2014mra,Zou:2018mxc}. The fitting is considered optimal when the value of ${{{\chi }^{2}}}/{\text{dof}}$ is closest to $1$. Consequently, we fix the endpoint and vary the starting point to extract the mass. The TCDC and the best-fitting curve for the ensemble $24^{3}\times48$ with $\tau=0.34$ (or $\sqrt{8\tau}=0.15\thinspace\text{fm}$) as an example are illustrated in Fig. \ref{C_r-vs-r}.

\begin{figure}[H]
\includegraphics[scale=0.73]{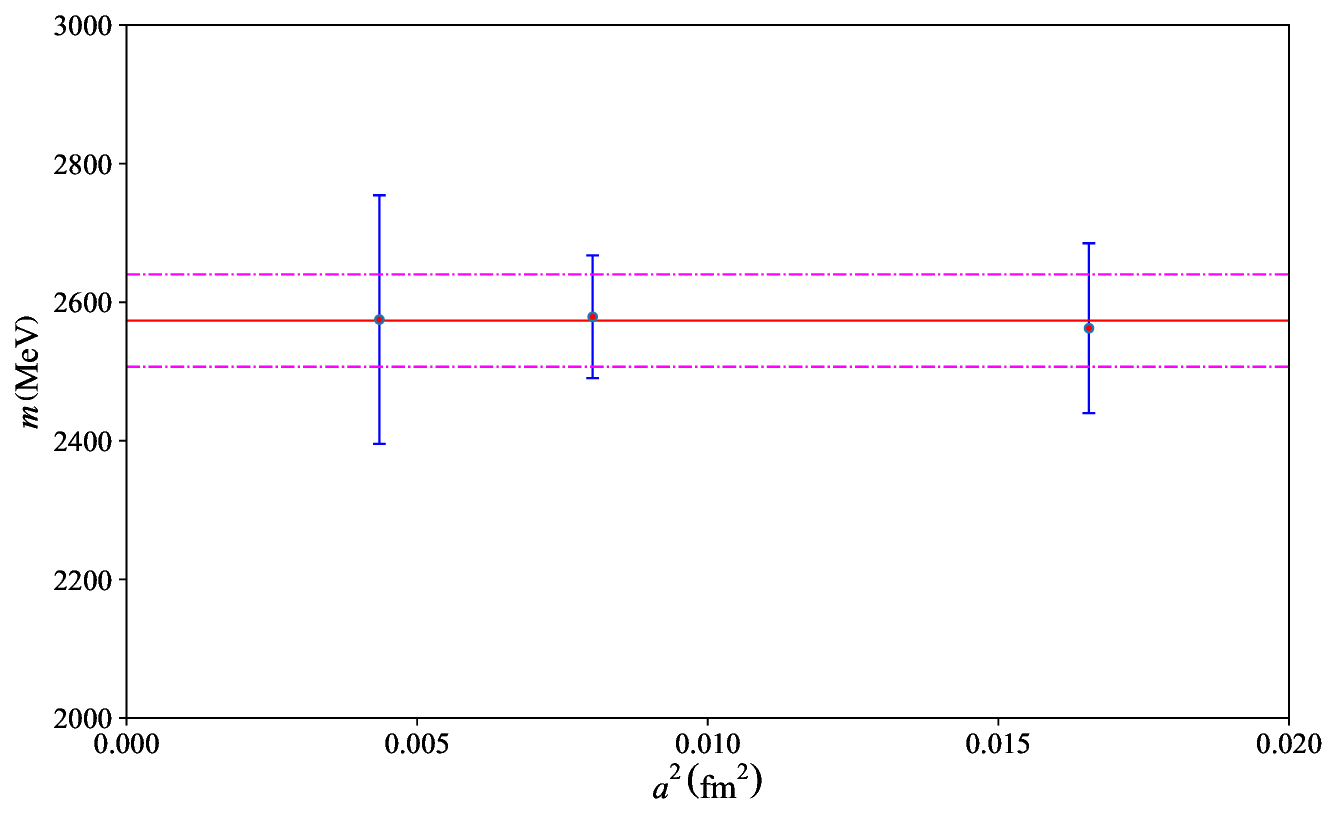}
\caption{ The colored scatter points represent the particle masses obtained from curve fitting across different lattice ensembles. The red solid line shows the continuum value of the pseudoscalar glueball mass, and the magenta lines indicate the errors.}

\label{mass-extra}
\end{figure}

The optimal fit results for the three ensembles reveal minimal differences, as shown in Tab. \ref{mass-res}. To obtain the particle mass via continuous extrapolation, we perform a constant fit. The plot of mass $M$ versus $a^{2}$, along with the fitting results, is illustrated in Fig. \ref{mass-extra}. The red solid line represents the mass of the pseudoscalar glueball, while the magenta lines indicate the associated errors. The mass of the pseudoscalar glueball, obtained through continuous extrapolation, is $m=2573\left(67\right)\text{MeV}$, consistent with the findings in Ref.  \cite{Chowdhury:2014mra,Chen:2005mg,Athenodorou:2020ani}.

\begin{table}[H]
	\begin{spacing}{1.5}
		\begin{centering}
			\begin{tabular*}{15cm}{@{\extracolsep{\fill}}ccccc}
\hline 
\makecell[c]{$L^{3}\times T$} & \makecell[c]{\small{Fitting interval$\thinspace\left(\text{fm}\right)$}} & \makecell[c]{$am$} & \makecell[c]{$m\thinspace\left(\text{MeV}\right)$} &  \makecell[c]{$\chi^{2}/{\text{dof}}$} \tabularnewline
\hline 
\makecell[c]{$16^{4}$} & \makecell[c]{$\left[0.498,0.863\right]$} & \makecell[c]{$1.682\left(81\right)$} & \makecell[c]{$2562\left(123\right)$} & \makecell[c]{$1.023$} \tabularnewline
\makecell[c]{$24^{3}\times48$} & \makecell[c]{$\left[0.40,0.627\right]$} & \makecell[c]{$1.162\left(40\right)$} & \makecell[c]{$2579\left(89\right)$} & \makecell[c]{$1.198$} \tabularnewline
\makecell[c]{$32^{4}$} & \makecell[c]{$\left[0.390,0.507\right]$} & \makecell[c]{$0.861\left(60\right)$} & \makecell[c]{$2575\left(179\right)$} & \makecell[c]{$1.198$} \tabularnewline
\hline 
\end{tabular*}
\par\end{centering}
\end{spacing}
\caption{ The fitting detailed parameters from TCDC for different ensembles. }

\label{mass-res}
\end{table}

\section{Conclusions}

The topological charge density of lattice QCD using both the fermionic and gluonic definitions is analyzed in this paper. The topological charge density for the fermionic definition is calculated using the SMP method, while the gluonic definition employs the Wilson flow. The SMP method offers advantages in computing the topological charge for the fermionic definition compared to the point sources, particularly with the parameters $(4,2)$, which is used effectively to compute the topological charge of the fermionic definition and significantly reduce computational resource requirements. The SMP method with parameters $(4,2)$ offers a potential approach for calculating TCDC and the topological susceptibility of the fermionic definition. The SMP method with parameters $(4,2)$ can effectively determine the matching Wilson flow time, $\tau_{\text{mr}}$, for the TCDC of the gluonic definition, as well as $\tau_{\text{qmr}}$ for the topological charge defined by the gluonic field. Studies have shown that the matching flow time can be determined using the SMP method, providing a benchmark for selecting the Wilson flow time for calculating the topological charge and its density correlation function. However, the universality of this method requires further investigation. 

The TCDC of the gluonic definition has also been analyzed using the Wilson flow in this work. Given its severe singularities and lattice artifacts, the Wilson flow serves as an effective smoothing method. We employ the TCDC calculated at the matching Wilson flow time to extract the pseudoscalar glueball mass through curve fitting. The pseudoscalar glueball mass obtained from the continuous extrapolation of three ensembles is consistent with results from other studies.

In future work, we will condcut a systematic comparison of various topological quantities obtained using the SMP source vectors and the point source vectors to quantify the trade-off between accuracy and computational cost. More larger ensemble will be required to validate the stability of the matching flow time across configurations and to assess the robustness of the SMP method against statistical fluctuations. We will attempt to use the SMP method with parameters $(4,2)$ to compute the topological charge density, topological charge, and TCDC of fermionic definition with a large number of configurations, and to conduct a detailed investigation of the properties of these topological quantities. Furthermore, based on this study, we will carry out a more comprehensive analysis of the properties of the matching Wilson flow time and explore its potential applications. We may also consider improving the method for determining the matching Wilson flow time to enhance the accuracy and effectiveness of the calculations, and research the effect of the smearing radius. 

\begin{centering}
\acknowledgments
\end{centering}

We thank Jian-bo Zhang and Yi-bo Yang for useful discussions and suggestions. Most Numerical simulations have been performed on the Tianhe-$2$ supercomputer at the National Supercomputer Center in Guangzhou (NSCC-GZ), China. This research was supported by the National Natural Science Foundation of China (NSFC) under the project No. $\text{11335001}$ and Zhejiang Provincial Natural Science Foundation of China under Grant No. $\text{LQ23A050001}$.

\
\

\begin{centering}
\appendix
\section{Supplementary data results of the main text. \label{app-A}}
\end{centering}

\begin{figure}[htbp!]
	\noindent \begin{centering}
		\includegraphics[scale=0.64]{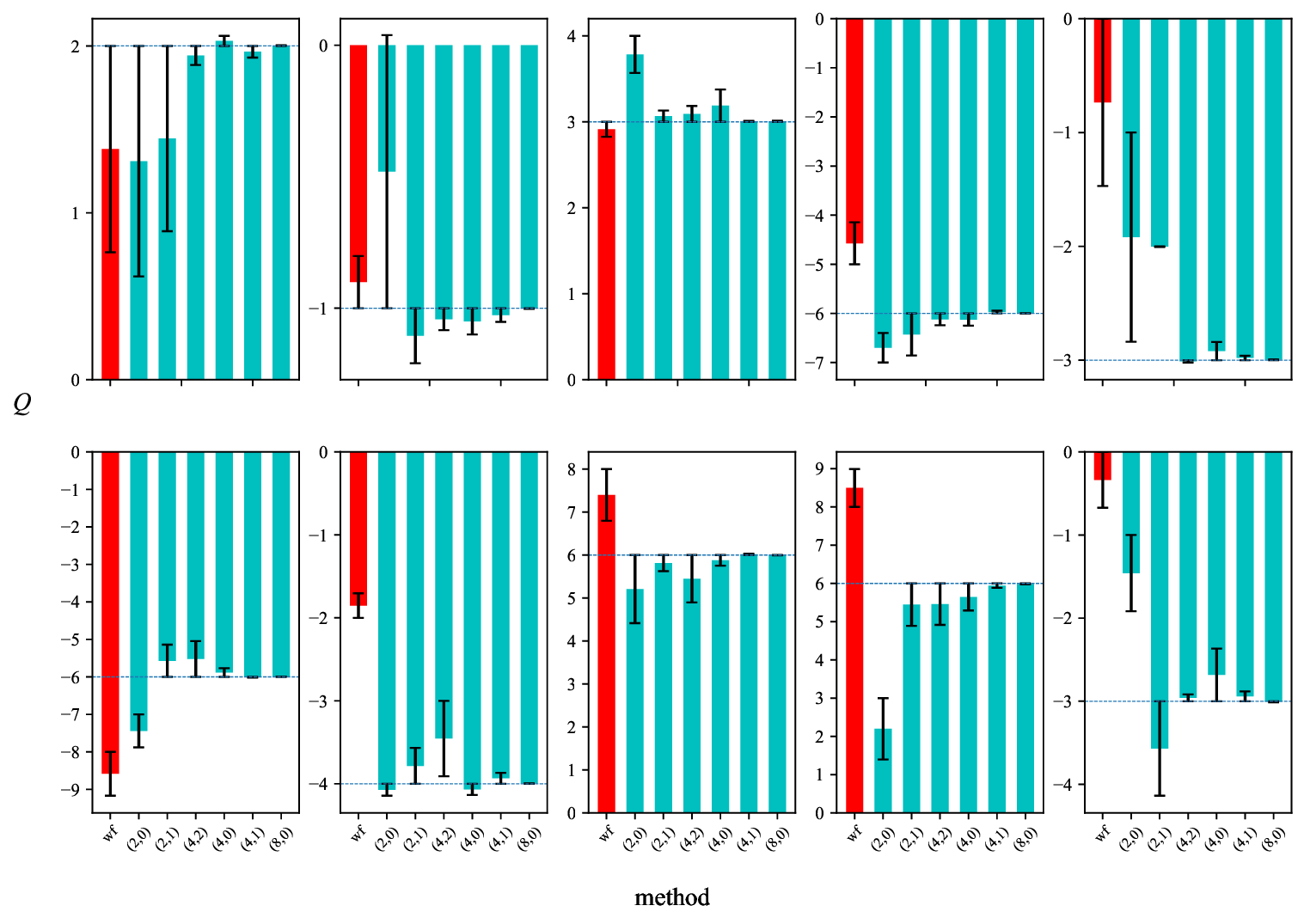}
		\par\end{centering}
	\caption{ The topological charge of the configurations calculated using the fermionic definition by the SMP method and the gluonic definition at the matching Wilson flow time $\tau_{\text{mr}}$. wf stands for the Wilson flow at $\tau_{\text{mr}}$. The error bars do not represent the uncertainties. Instead, they indicate the deviations of the topological charges $Q_{\text{smp}}$ calculated using the SMP method with different parameters $(d,\text{mode})$ and $Q_{\text{wf}}$ using the Wilson flow at the matching Wilson flow time from the topological charge $Q$. The integer topological charges $Q$ (the dashed line in the figure) are obtained by rounding the fermionic topological charges computed using the SMP method with parameters (8,0). The results for the lattices $16^{4}$ and $24^{3}\times48$ are presented from top to bottom, and the configurations for each lattice are shown from left to right.}
	\label{Q-diff-2}
\end{figure}

In this appendix, the data and figures that are not included in the main text are shown. Topological charges of five additional configurations calculated using the SMP method at $\kappa = 0.18$ and the Wilson flow at $\tau_{\text{mr}}$ for the lattice ensembles $16^4$ and $24^3\times48$ are shown in Fig. \ref{Q-diff-2}, and the results are similar to those shown in Fig. \ref{Q-diff-1}.

The topological charges of the fermionic definition calculated by the SMP method with $\kappa = 0.19$, the topological charges calculated using the gluonic definition by the Wilson flow, and the matching Wilson flow time $\tau_{\text{qmr}}$ are presented in Tab. \ref{Q-tao-pr-3} and \ref{Q-tao-pr-4}. The results for $\kappa = 0.19$ are basically consistent with those for $\kappa = 0.18$, leading to the same conclusion.

\begin{table}[H]
	\begin{tabular}{|m{1.9cm}|m{1.9cm}|m{1.9cm}|m{1.9cm}|m{1.9cm}|m{1.9cm}|m{1.9cm}|m{1.9cm}|}
		\hline 
		\makecell[c]{$L^3\times T$} & \makecell[c]{S/N} & \makecell[c]{$Q$} & \makecell[c]{$Q_{\text{smp}}$} & \makecell[c]{$\left|Q-Q_{\text{smp}}\right|$} & \makecell[c]{$Q_{\text{wf}}$} & \makecell[c]{$\left|Q-Q_{\text{wf}}\right|$} & \makecell[c]{$\tau_{\text{qmr}}$} \tabularnewline
		\hline 
		\makecell[c]{\multirow{5}{*}{$16^{4}$}} & \makecell[c]{$10$} & \makecell[c]{$2$} & \makecell[c]{$1.9999$} & \makecell[c]{$0.0001$} & \makecell[c]{$2.6857$} & \makecell[c]{$0.6857$} & \makecell[c]{$0.64$} \tabularnewline
		\cline{2-8} \cline{3-8} \cline{4-8} \cline{5-8} \cline{6-8} \cline{7-8} \cline{8-8} 
		& \makecell[c]{$18$} & \makecell[c]{$7$} & \makecell[c]{$7.0020$} & \makecell[c]{$0.0020$}  & \makecell[c]{$6.8058$} & \makecell[c]{$0.1942$} & \makecell[c]{$1.00$} \tabularnewline
		\cline{2-8} \cline{3-8} \cline{4-8} \cline{5-8} \cline{6-8} \cline{7-8} \cline{8-8}
		& \makecell[c]{$20$} & \makecell[c]{$3$} & \makecell[c]{$3.0028$} & \makecell[c]{$0.0028$}  & \makecell[c]{$3.0365$} & \makecell[c]{$0.0365$} & \makecell[c]{$0.40$} \tabularnewline
		\cline{2-8} \cline{3-8} \cline{4-8} \cline{5-8} \cline{6-8} \cline{7-8} \cline{8-8}
		& \makecell[c]{$22$} & \makecell[c]{$-1$} & \makecell[c]{$-0.9993$} & \makecell[c]{$0.0007$}  & \makecell[c]{$-1.0952$} & \makecell[c]{$0.09516$} & \makecell[c]{$0.90$} \tabularnewline
		\cline{2-8} \cline{3-8} \cline{4-8} \cline{5-8} \cline{6-8} \cline{7-8} \cline{8-8} 
		& \makecell[c]{$30$} & \makecell[c]{$-4$} & \makecell[c]{$-4.0028$} & \makecell[c]{$0.0028$}  & \makecell[c]{$-4.0206$}  & \makecell[c]{$0.0206$} & \makecell[c]{$0.38$} \tabularnewline
		\hline 
		\makecell[c]{\multirow{5}{*}{$24^{3}\times48$}} & \makecell[c]{$5$} & \makecell[c]{$3$} & \makecell[c]{$3.0055$} & \makecell[c]{$0.0055$} & \makecell[c]{$2.9521$}  & \makecell[c]{$0.0479$}  & \makecell[c]{$1.00$} \tabularnewline
		\cline{2-8} \cline{3-8} \cline{4-8} \cline{5-8} \cline{6-8} \cline{7-8} \cline{8-8} 
		& \makecell[c]{$15$} & \makecell[c]{$-3$} & \makecell[c]{$-3.0087$} & \makecell[c]{$0.0087$}  & \makecell[c]{$-2.9871$}  & \makecell[c]{$0.0129$}  & \makecell[c]{$0.26$} \tabularnewline
		\cline{2-8} \cline{3-8} \cline{4-8} \cline{5-8} \cline{6-8} \cline{7-8} \cline{8-8} 
		& \makecell[c]{$17$} & \makecell[c]{$6$} & \makecell[c]{$5.9922$} & \makecell[c]{$0.0078$}  & \makecell[c]{$5.9620$}  & \makecell[c]{$0.03799$}  & \makecell[c]{$0.34$} \tabularnewline
		\cline{2-8} \cline{3-8} \cline{4-8} \cline{5-8} \cline{6-8} \cline{7-8} \cline{8-8} 
		& \makecell[c]{$30$} & \makecell[c]{$-5$} & \makecell[c]{$-4.9897$}  & \makecell[c]{$0.0103$}  & \makecell[c]{$-4.9756$} & \makecell[c]{$0.0244$}  & \makecell[c]{$0.44$} \tabularnewline
		\cline{2-8} \cline{3-8} \cline{4-8} \cline{5-8} \cline{6-8} \cline{7-8} \cline{8-8} 
		& \makecell[c]{$53$} & \makecell[c]{$4$} & \makecell[c]{$4.0074$} & \makecell[c]{$0.0074$}  & \makecell[c]{$4.0062$}  & \makecell[c]{$0.0062$}  & \makecell[c]{$0.26$} \tabularnewline 
		\hline 
	\end{tabular}
	
	\caption{ The matching Wilson flow time $\tau_{\text{qmr}}$ for the topological charge $Q_{\text{wf}}$ calculated using the Wilson flow. $Q$ is obtained by rounding the result $Q_{\text{smp}}$ of the SMP method with parameters $(8,0)$ to the nearest integer and $\kappa = 0.19$. S/N is the sequence number of the configuration. \label{Q-tao-pr-3}}
\end{table}

\begin{table}[H]
	\begin{tabular}{|m{1.9cm}|m{1.9cm}|m{1.9cm}|m{1.9cm}|m{1.9cm}|m{1.9cm}|m{1.9cm}|m{1.9cm}|}
		\hline 
		\makecell[c]{$L^3\times T$} & \makecell[c]{S/N} & \makecell[c]{$Q$} & \makecell[c]{$Q_{\text{smp}}$} & \makecell[c]{$\left|Q-Q_{\text{smp}}\right|$} & \makecell[c]{$Q_{\text{wf}}$} & \makecell[c]{$\left|Q-Q_{\text{wf}}\right|$} & \makecell[c]{$\tau_{\text{qmr}}$} \tabularnewline
		\hline 
		\makecell[c]{\multirow{5}{*}{$16^{4}$}} & \makecell[c]{$10$} & \makecell[c]{$3$} & \makecell[c]{$1.8717$} & \makecell[c]{$0.1283$} & \makecell[c]{$2.6857$} & \makecell[c]{$0.6857$} & \makecell[c]{$0.64$} \tabularnewline
		\cline{2-8} \cline{3-8} \cline{4-8} \cline{5-8} \cline{6-8} \cline{7-8} \cline{8-8} 
		& \makecell[c]{$18$} & \makecell[c]{$7$} & \makecell[c]{$6.8934$} & \makecell[c]{$0.1066$} & \makecell[c]{$6.8058$} & \makecell[c]{$0.1942$} & \makecell[c]{$1.00$} \tabularnewline
		\cline{2-8} \cline{3-8} \cline{4-8} \cline{5-8} \cline{6-8} \cline{7-8} \cline{8-8} 
		& \makecell[c]{$20$} & \makecell[c]{$3$} & \makecell[c]{$2.8148$} & \makecell[c]{$0.1852$} & \makecell[c]{$3.0365$} & \makecell[c]{$0.0365$} & \makecell[c]{$0.40$} \tabularnewline
		\cline{2-8} \cline{3-8} \cline{4-8} \cline{5-8} \cline{6-8} \cline{7-8} \cline{8-8}
		& \makecell[c]{$22$} & \makecell[c]{$-1$} & \makecell[c]{$-0.9084$} & \makecell[c]{$0.0916$} & \makecell[c]{$-1.0952$} & \makecell[c]{$0.0952$} & \makecell[c]{$0.90$} \tabularnewline
		\cline{2-8} \cline{3-8} \cline{4-8} \cline{5-8} \cline{6-8} \cline{7-8} \cline{8-8} 
		& \makecell[c]{$30$} & \makecell[c]{$-4$} & \makecell[c]{$-4.2922$} & \makecell[c]{$0.2922$} & \makecell[c]{$-4.0206$} & \makecell[c]{$0.0206$} & \makecell[c]{$0.38$} \tabularnewline
		\hline 
		\makecell[c]{\multirow{5}{*}{$24^{3}\times48$}} & \makecell[c]{$5$} & \makecell[c]{$3$} & \makecell[c]{$2.8990$} & \makecell[c]{$0.1010$} & \makecell[c]{$2.9521$} & \makecell[c]{$0.0479$} & \makecell[c]{$1.00$} \tabularnewline
		\cline{2-8} \cline{3-8} \cline{4-8} \cline{5-8} \cline{6-8} \cline{7-8} \cline{8-8} 
		& \makecell[c]{$15$} & \makecell[c]{$-3$} & \makecell[c]{$-1.9533$} & \makecell[c]{$1.0467$} & \makecell[c]{$-2.1213$} & \makecell[c]{$0.8787$} & \makecell[c]{$0.18$} \tabularnewline
		\cline{2-8} \cline{3-8} \cline{4-8} \cline{5-8} \cline{6-8} \cline{7-8} \cline{8-8} 
		& \makecell[c]{$17$} & \makecell[c]{$6$} & \makecell[c]{$5.9715$} & \makecell[c]{$0.0285$} & \makecell[c]{$5.9620$} & \makecell[c]{$0.0380$} & \makecell[c]{$0.34$} \tabularnewline
		\cline{2-8} \cline{3-8} \cline{4-8} \cline{5-8} \cline{6-8} \cline{7-8} \cline{8-8}
		& \makecell[c]{$30$} & \makecell[c]{$-5$} & \makecell[c]{$-5.3063$} & \makecell[c]{$0.3063$} & \makecell[c]{$-4.9756$} & \makecell[c]{$0.0244$} & \makecell[c]{$0.44$}\tabularnewline
		\cline{2-8} \cline{3-8} \cline{4-8} \cline{5-8} \cline{6-8} \cline{7-8} \cline{8-8}
		& \makecell[c]{$53$} & \makecell[c]{$3$} & \makecell[c]{$3.3449$} & \makecell[c]{$0.6551$} & \makecell[c]{$3.7656$} & \makecell[c]{$0.2344$} & \makecell[c]{$0.18$}\tabularnewline
		\hline 
	\end{tabular}
	
	\caption{ The matching Wilson flow time $\tau_{\text{qmr}}$ for calculating $Q$ of the gluonic definition. $Q$ is obtained by rounding the result of the SMP method with parameters $(4,2)$ to the nearest integer and $\kappa = 0.19$. S/N stands for the sequence number of configurations. \label{Q-tao-pr-4}}
\end{table}

The matching parameters of the SMP method, obtained by comparing $(d \ne 8, \text{mode} \ne 0)$ with $(8,0)$, for the remaining five representative configurations on the $16^4$ and $24^3\times48$ lattices at $\kappa = 0.18$ and $\kappa = 0.19$, are presented in Tab. \ref{Q-tao-pr-3} and \ref{Q-tao-pr-4}, respectively. The results exhibit the same conclusion as those presented in the Tab \ref{Q-tao-pr-1} and \ref{Q-tao-pr-2}.

\begin{table}[htbp!]
	\renewcommand{\arraystretch}{1.2} 
		\begin{centering}
			\begin{tabular}{|m{2.0cm}<{\centering}|m{2.2cm}<{\centering}|m{2.2cm}<{\centering}|m{2.2cm}<{\centering}|m{2.2cm}<{\centering}|m{2.2cm}<{\centering}|m{2.2cm}<{\centering}|}
				\hline 
				$\text{$L^3\times T$}$ & S/N & $\left(2,0\right)$ & $\left(2,1\right)$ & $\left(4,2\right)$ & $\left(4,0\right)$ & $\left(4,1\right)$\\ \hline
				
				\multirow{10}{*}{\centering $16^{4}$} 
				& \multirow{2}{*}{54} 
				& 0.626647 & 0.811648 & 0.976293 & 0.987963 & 0.999219 \\
				& & (10946) & (13305) & (15525) & (15671) & (15828) \\ \cline{2-7}
				& \multirow{2}{*}{78} 
				& 0.637919 & 0.817856 & 0.976464 & 0.988026 & 0.999229 \\
				& & (11811) & (14443) & (16737) & (16910) & (17081) \\ \cline{2-7}
				& \multirow{2}{*}{85}  
				& 0.631974 & 0.813821 & 0.975856 & 0.987865 & 0.999216 \\
				& & (11443) & (13961) & (16313) & (16511) & (16673) \\ \cline{2-7}
				& \multirow{2}{*}{88}  
				& 0.636668 & 0.815576 & 0.976374 & 0.988077 & 0.999236 \\
				& & (11299) & (13843) & (16173) & (16328) & (16484) \\ \cline{2-7}
				& \multirow{2}{*}{96}  
				& 0.622198 & 0.810164 & 0.975371 & 0.987558 & 0.999214 \\ 
				& & (10728) & (13118) & (15313) & (15485) & (15626) \\ \hline
				
				\multirow{10}{*}{\centering $24^{3}\times48$} 
				& \multirow{2}{*}{54} 
				& 0.619361 & 0.808776 & 0.977361 & 0.988133 & 0.999402 \\
				& & (3261) & (4028) & (4719) & (4765) & (4814) \\ \cline{2-7}
				& \multirow{2}{*}{82} 
				& 0.617199 & 0.807676 & 0.977178 & 0.988004 & 0.999397 \\
				& & (3264) & (4035) & (4743) & (4785) & (4831) \\ \cline{2-7}
				& \multirow{2}{*}{91} 
				& 0.618554 & 0.807475 & 0.977126 & 0.988019 & 0.999388 \\
				& & (3263) & (4042) & (4740) & (4785) & (4831) \\ \cline{2-7}
				& \multirow{2}{*}{93} 
				& 0.619698 & 0.809922 & 0.977506 & 0.988176 & 0.999400 \\
				& & (3323) & (4125) & (4826) & (4870) & (4920) \\ \cline{2-7}
				& \multirow{2}{*}{95} 
				& 0.616591 & 0.807393 & 0.977036 & 0.987982 & 0.999389 \\
				& & (3234) & (4001) & (4691) & (4735) & (4781) \\ \hline				
			\end{tabular}
			\par\end{centering}
		\caption{$\text{\ensuremath{\Xi_{AB}}}$ for SMP method, obtained by comparing different parameters $(d\ne8,\text{mode}\ne0)$ to $\left(8,0\right)$. The hopping parameter
			is $\kappa=0.18$. S/N stands for the sequence number of configurations. The values in parentheses indicate the standard deviation. To distinguish the computed results, six significant digits were used.}
		\label{xi_AB_app_1}
\end{table}

\begin{table}[htbp!]
	\renewcommand{\arraystretch}{1.2} 
		\begin{centering}
			\begin{tabular}{|m{2.0cm}<{\centering}|m{2.2cm}<{\centering}|m{2.2cm}<{\centering}|m{2.2cm}<{\centering}|m{2.2cm}<{\centering}|m{2.2cm}<{\centering}|m{2.2cm}<{\centering}|}
				\hline 
				$\text{$L^3\times T$}$ & S/N & $\left(2,0\right)$ & $\left(2,1\right)$ & $\left(4,2\right)$ & $\left(4,0\right)$ & $\left(4,1\right)$\\ \hline
				
				\multirow{10}{*}{\centering $16^{4}$} 
				& \multirow{2}{*}{54} 
				& 0.667060 & 0.844807 & 0.983191 & 0.990720 & 0.999546 \\
				& & (11623) & (13925) & (15859) & (15961) & (16094) \\ \cline{2-7}
				& \multirow{2}{*}{78} 
				& 0.678401 & 0.850165 & 0.983555 & 0.990851 & 0.999564 \\
				& & (12612) & (15182) & (17161) & (17272) & (17401) \\ \cline{2-7}
				& \multirow{2}{*}{85}  
				& 0.675642 & 0.848065 & 0.983276 & 0.990826 & 0.999558 \\
				& & (12707) & (15234) & (17349) & (17489) & (17627) \\ \cline{2-7}
				& \multirow{2}{*}{88}  
				& 0.677014 & 0.848266 & 0.983546 & 0.990902 & 0.999562 \\
				& & (12100) & (14597) & (16634) & (16727) & (16858) \\ \cline{2-7}
				& \multirow{2}{*}{96}  
				& 0.664842 & 0.844171 & 0.982761 & 0.990445 & 0.999545 \\ 
				& & (12002) & (14582) & (16604) & (16734) & (16860) \\ \hline
				
				\multirow{10}{*}{\centering $24^{3}\times48$} 
				& \multirow{2}{*}{54} 
				& 0.646677 & 0.831640 & 0.981828 & 0.990112 & 0.999551 \\
				& & (3392) & (4151) & (4779) & (4815) & (4856) \\ \cline{2-7}
				& \multirow{2}{*}{82} 
				& 0.645377 & 0.830907 & 0.981717 & 0.990006 & 0.999548 \\
				& & (3404) & (4163) & (4797) & (4830) & (4869) \\ \cline{2-7}
				& \multirow{2}{*}{91} 
				& 0.647505 & 0.831375 & 0.981829 & 0.990088 & 0.999547 \\
				& & (3489) & (4323) & (4987) & (5026) & (5065) \\ \cline{2-7}
				& \multirow{2}{*}{93} 
				& 0.647641 & 0.832914 & 0.982036 & 0.990171 & 0.999551 \\
				& & (3455) & (4240) & (4868) & (4901) & (4943) \\ \cline{2-7}
				& \multirow{2}{*}{95} 
				& 0.644754 & 0.830645 & 0.981657 & 0.989987 & 0.999543 \\
				& & (3371) & (4127) & (4745) & (4779) & (4817) \\ \hline		
			\end{tabular}
			\par\end{centering}
		\caption{$\text{\ensuremath{\Xi_{AB}}}$ for SMP method, determined by comparing different parameters $(d\ne8,\text{mode}\ne0)$ to $\left(8,0\right)$. The hopping parameter
			is $\kappa=0.19$. S/N stands for the sequence number of configurations. The values in parentheses indicate the standard deviation. To distinguish the computed results, six significant digits were used.}
		\label{xi_AB_app_2}
\end{table}

\begin{figure}[H]
	\noindent \begin{centering}
		\includegraphics[scale=0.54]{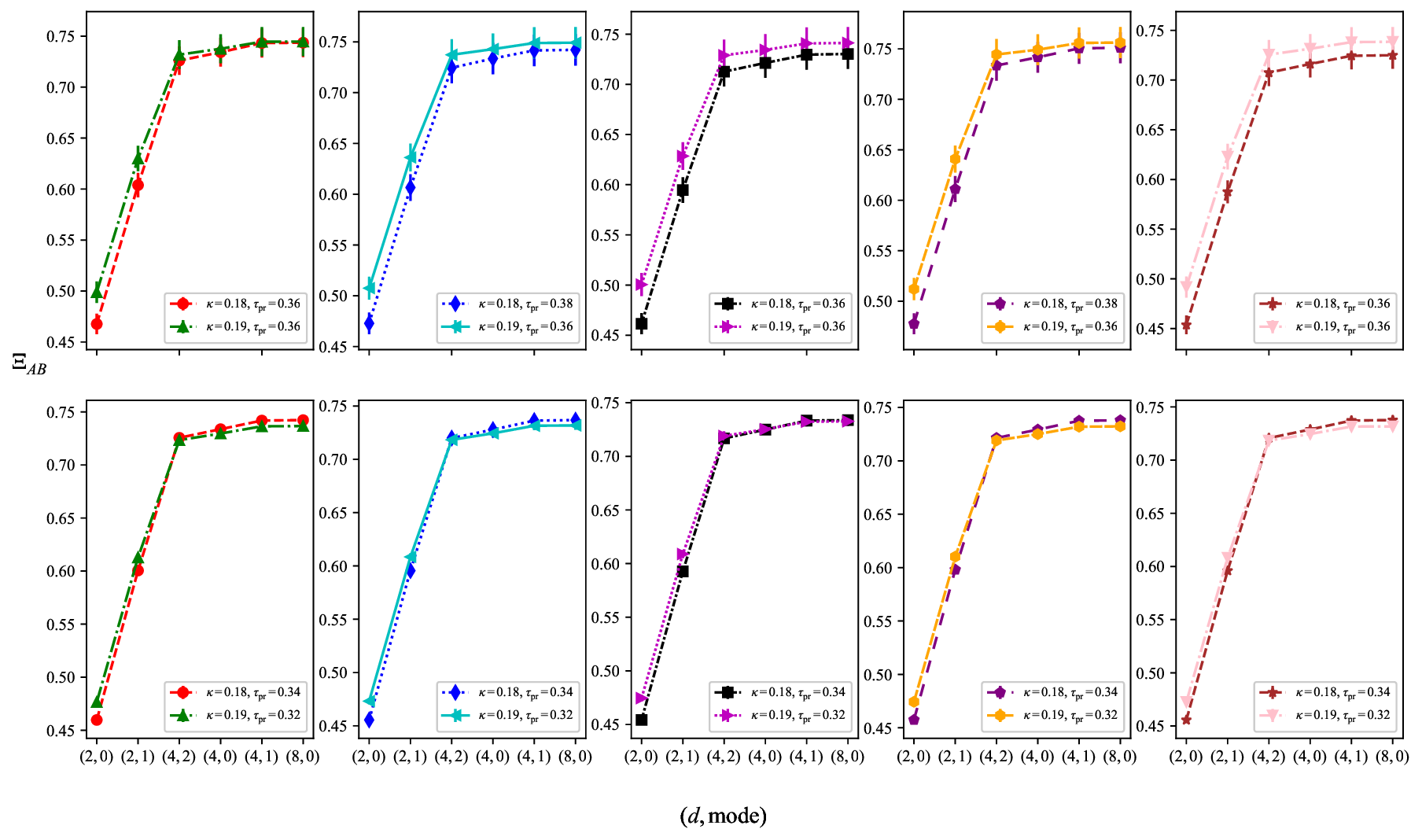}
		\par\end{centering}
	\caption{ The best $\text{\ensuremath{\Xi_{AB}}}$ for SMP with different parameters $\left(d,\text{mode}\right)$ compared with the Wilson flow in the calculation of topological charge density. $\tau_{\text{mr}}$ is the matching Wilson flow time, and $\kappa$ is the hopping parameter. From top to bottom, the results for the lattices $16^{4}$ and $24^{3}\times48$ are presented. The results of different configurations for each lattice are shown from left to right.}
	\label{xi_diff-2}
\end{figure}

For the lattice ensembles $16^4$ and $24^3\times48$, the best  $\text{\ensuremath{\Xi_{AB}}}$ for SMP with different parameters $\left(d,\text{mode}\right)$, compared with the Wilson flow in the calculation of the topological charge density, are shown in Fig. \ref{xi_diff-2}, and exhibits features similar to those shown in Fig.  \ref{xi_diff-1}.

\normalem

\providecommand{\href}[2]{#2}\begingroup\raggedright\endgroup

\end{document}